\documentclass[useAMS,usenatbib,usegraphicx,color]{mn2e}

\usepackage{color}
\usepackage{graphics}
\usepackage{psfig}
\usepackage{epsf}

\title[Circumstellar features in hot DA white dwarfs]
  {Circumstellar features in hot DA white dwarfs}
\author[N.P.Bannister, M.A.Barstow, J.B.Holberg, F.C.Bruhweiler]
  {N.P.~Bannister\thanks{Email: npb@star.le.ac.uk}$^1$, M.A.~Barstow$^1$, J.B.Holberg$^2$, F.C.Bruhweiler$^3$ \\
  $^1$Department of Physics and Astronomy, University of
Leicester, University Road, Leicester, LE1 7RH, U.K.\\ $^2$Lunar
and Planetary Laboratory, University of Arizona, Tucson, AZ 85721,
USA\\ $^3$Institute for Astrophysics \& Computational Sciences
(IACS), Department of Physics,\\ The Catholic University of
America, Washington DC 20064, USA}
\date{Accepted YYYY MMMMM DD. Received YYYY MMMMM DD}
\pagerange{\pageref{firstpage}--\pageref{lastpage}} \pubyear{1994}

\usepackage{natbib}
\usepackage{graphics}

\def\pcm3{{\rm\thinspace cm^{-3}}}

\def\contcaption{\@conttrue\SFB@caption\@captype}

\newcommand{\hi}{\mbox{H{\small\hspace{0.5mm}I}}}

\newcommand{\heii}{\mbox{He{\small\hspace{0.5mm}II}}}

\newcommand{\niv}{\mbox{Ni{\small\hspace{0.5mm}V}}}
\newcommand{\alii}{\mbox{Al{\small\hspace{0.5mm}II}}}
\newcommand{\aliii}{\mbox{Al{\small\hspace{0.5mm}III}}}
\newcommand{\civ}{\mbox{C{\small\hspace{0.5mm}IV}}}

\newcommand{\cii}{\mbox{C{\small\hspace{0.5mm}II}}}
\newcommand{\siiv}{\mbox{Si{\small\hspace{0.5mm}IV}}}
\newcommand{\siiii}{\mbox{Si{\small\hspace{0.5mm}III}}}
\newcommand{\siii}{\mbox{Si{\small\hspace{0.5mm}II}}}
\newcommand{\SII}{\mbox{S{\small\hspace{0.5mm}II}}}
\newcommand{\mgii}{\mbox{Mg{\small\hspace{0.5mm}II}}}

\newcommand{\oi}{\mbox{O{\small\hspace{0.5mm}I}}}
\newcommand{\oiv}{\mbox{O{\small\hspace{0.5mm}IV}}}
\newcommand{\oiii}{\mbox{O{\small\hspace{0.5mm}III}}}
\newcommand{\ov}{\mbox{O{\small\hspace{0.5mm}V}}}

\newcommand{\nv}{\mbox{N{\small\hspace{0.5mm}V}}}
\newcommand{\NI}{\mbox{N{\small\hspace{0.5mm}I}}}
\newcommand{\NII}{\mbox{N{\small\hspace{0.5mm}II}}}
\newcommand{\fev}{\mbox{Fe{\small\hspace{0.5mm}V}}}
\newcommand{\feiii}{\mbox{Fe{\small\hspace{0.5mm}III}}}
\newcommand{\pv}{\mbox{P{\small\hspace{0.5mm}V}}}

\def\n_h{{\rm n_{H}}}

\def\syn{{\small SYNSPEC}}
\def\tlu{{\small TLUSTY}}
\def\err{$\pm$}
\def\kms{{km\thinspace s$^{-1}$}}
\def\deg{$^{\circ}$}
\def\lam{$\lambda$}

\def\G1{{\sc G191-B2B}}

\def\tl{{\sc TLUSTY}}

\def\EUVE{{\sl EUVE}}

\def\iue{{\sl IUE}}
\def\stis{{\sl STIS}}
\def\ghrs{{\sl GHRS}}

\def\per{{$^{-1}$}}
\def\sn{{S/N}}
\def\teff{{T$_{\hbox{\footnotesize{eff}}}$ }}
\def\mv{{m$_v$}}
\def\vgr{{$v_{\hbox{\tiny grav}}$}}
\def\vexp{{$v_{exp}$}}

\def\lg{{log {\it g} }}

\def\NH1{{$N_{\rm HI}~$}}

\def\ga{{\rm\thinspace gauss}}

\def\Rsun{\hbox{$\rm\thinspace R_{\odot}$}}
\def\lii{{\it l}$_{\hbox{\tiny II}}$}
\def\bii{{\it b}$_{\hbox{\tiny II}}$}
\def\ra{$\alpha$}
\def\dec{$\delta$}
\def\vlic{$v_{\hbox{\tiny LIC}}$}
\def\vgr{$v_{\hbox{\tiny grav}}$}
\def\vesc{$v_{\hbox{\tiny esc}}$}
\def\vlics{$v_{\hbox{\tiny LIC*}}$}
\def\vism{$v_{\hbox{\tiny ISM}}$}
\def\vphot{$v_{\hbox{\tiny phot}}$}
\def\vcirc{$v_{\hbox{\tiny circ}}$}

\def\Msun{\hbox{$\rm\thinspace M_{\odot}$}}

\def\approxlt{\mathrel{\hbox{\rlap{\lower .5ex \hbox {$\sim$}}
    \raise .15 ex \hbox{$<$}}}}
\def\approxgt{\mathrel{\hbox{\rlap{\lower .5ex \hbox {$\sim$}}
    \raise .15 ex \hbox{$>$}}}}

\def\la{\mathrel{\hbox{\rlap{\hbox{\lower4pt\hbox{$\sim$}}}\hbox{$<$}}}}
\def\ga{\mathrel{\hbox{\rlap{\hbox{\lower4pt\hbox{$\sim$}}}\hbox{$>$}}}}
\newbox\grsign \setbox\grsign=\hbox{$>$} \newdimen\grdimen
\grdimen=\ht\grsign
\newbox\simlessbox \newbox\simgreatbox \newbox\simpropbox
\setbox\simgreatbox=\hbox{\raise.5ex\hbox{$>$}\llap
     {\lower.5ex\hbox{$\sim$}}}\ht1=\grdimen\dp1=0pt
\setbox\simlessbox=\hbox{\raise.5ex\hbox{$<$}\llap
     {\lower.5ex\hbox{$\sim$}}}\ht2=\grdimen\dp2=0pt
\setbox\simpropbox=\hbox{\raise.5ex\hbox{$\propto$}\llap
     {\lower.5ex\hbox{$\sim$}}}\ht2=\grdimen\dp2=0pt

\begin{document}
\definecolor{gray}{rgb}{0.8,0.8,0.8}
\maketitle \label{firstpage}
\begin{abstract}
We present a phenomenological study of highly ionised,
non-photospheric absorption features in high spectral resolution
vacuum ultraviolet spectra of 23 hot DA white dwarfs. Prior to
this study, four of the survey objects (Feige 24, REJ 0457-281,
G191-B2B and REJ 1614-085) were known to possess these features.
We find four new objects with multiple components in one or more
of the principal resonance lines: REJ 1738+665, Ton 021, REJ
0558-373 and WD 2218+706. A fifth object, REJ 2156-546 also shows
some evidence of multiple components, though further observations
are required to confirm the detection. We discuss possible origins
for these features including ionisation of the local interstellar
environment, the presence of material inside the gravitational
well of the white dwarf, mass loss in a stellar wind, and the
existence of material in an ancient planetary nebula around the
star. We propose ionisation of the local interstellar medium as
the origin of these features in G191-B2B and REJ 1738+665, and
demonstrate the need for higher resolution spectroscopy of the
sample, to detect multiple ISM velocity components and to identify
circumstellar features which may lie close to the photospheric
velocity.

\end{abstract}


\section{Introduction}
Spectral lines from highly ionised species are observed at
non-photospheric velocities in several hot DO white dwarfs. Of the
11 stars included in the survey presented by~\citet{Hol:1998a}
(hereafter referred to as HBS), six objects, all with temperatures
\teff $\geq$ 70,000 K, were found to exhibit highly ionised
features at non-photospheric velocities. ~\citet{Hol:1999d} argue
that such features cannot be of an interstellar origin since such
highly ionised species are uncharacteristic of the local ISM, and
are not observed along adjacent lines of sight to stars at greater
distances. Further, while interstellar lines can be observed at
velocities which are blue- or red-shifted with respect to the
photospheric value \vphot, only blue-shifted features are found in
the DO sample discussed by HBS. Such features provide strong
evidence for the existence of ongoing mass loss in hot DO white
dwarfs.

Although over 80\%\ of known white dwarfs belong to the DA
group~\citep{Sio:1997}, relatively few have been found to exhibit
non-photospheric features. Only 5 of the 44 DA stars considered in
HBS are accompanied by any type of circumstellar feature, with no
apparent dependence on temperature. In some instances, the
appearance of this phenomenon can be plausibly explained in terms
of interactions between binary components, as in the DA + dM 1.5
system Feige 24, which shows multiple components in the \civ\ \lam
\lam~1548, 1550 doublet.

However, in the DA white dwarf CD- 38$^{\circ}$ 10980, which is a
member of a {\em wide binary} system,~\citet{Hol:1995b} find Si
and C absorption features in IUE spectra, which are shifted by -12
km s$^{-1}$ with respect to the photospheric velocity of the with
dwarf, which can be inferred from its measured gravitational
redshift.  Holberg et al. showed that the atmosphere of this
object was devoid of Si and C, at the expected photospheric
velocity, and used the presence of excited or meta-stable levels
in the shifted features as evidence that the material was not
located in the ISM along the line-of-sight to the star.  These
observations were explained in terms of a dense, gaseous halo in
close proximity to the star, possibly an extension to the
atmosphere, for which a similar temperature and electron density
was derived.  Alternately,~\citet{Wol:2001} using FUSE
observations of CD- 38$^{\circ}$ 10980, observed a set of Si III
lines shortward of 1120 \AA\ which they attribute to the stellar
photosphere.  They were able to successfully model the observed
equivalent widths of these lines as well as the Si lines seen in
the IUE observations with a photospheric Si abundance of
2x10$^{-8}$. However, the velocity discrepancy remains
unexplained. In another example of circumstellar features in an
isolated DA object,~\citet{Hol:1997b} found evidence for weakly
blue-shifted \civ\ and \siiv\ components in REJ 1614-085 (\teff
$\sim$ 38,500 K) at $\approx$ 30\%\ of the strength of the
photospheric lines, shifted by -25 and -40 \kms\ respectively.
Similar features in the spectrum of the \teff\ $\sim$ 57,000 K DA,
REJ 0457-289 have also been discussed by~\citet{Hol:1997a}.

Agreement between the predicted and observed abundances of atomic
species in the atmospheres of white dwarf stars has improved with
the introduction of stratified model atmosphere codes, beginning
with the stratification of He and Fe investigated
by~\citet{Bar:1998b} and~\citet{Bar:1999a}.~\citet{Dre:2001}
 describe models in which stratified
abundances are calculated self-consistently, by considering the
depth dependence of temperature, density, radiation field and
level populations using an iterative procedure. These models have
succeeded in reproducing the soft X-ray, EUV and FUV spectra of a
sample of DA stars, and can explain the widely varying levels of
metallicity in hot DA white dwarfs. However, this agreement is not
complete; for example, lines of C, N and O are not reproduced as
accurately as those of Fe and Ni, while observed differences
between objects of similar \teff\ and \lg\ are
unexplained~\citep{Sch:2001}. The success of stratified models is
encouraging, but the observation of highly ionised circumstellar
features suggests the existence of processes which may modify the
predicted equilibrium abundances, and improving the agreement
between observed and predicted abundances requires better
understanding of the nature of these features. It is particularly
interesting that the DO white dwarfs, in which circumstellar
features are relatively common, are also poorly modeled by the new
generation of stratified codes~\citep{Dre:2001}.

Data from the \stis\ instrument on board HST have allowed the
study of white dwarf spectra to be carried out with an accuracy
impossible to achieve using earlier instruments, and at a
resolution which permits more precise examination of intrinsic
line profiles. A more sensitive search can now be made for signs
of mass loss and accretion which may modify equilibrium abundances
in white dwarf envelopes.

In section~\ref{sec:obs} we present a phenomenological study of a
sample of 23 hot (\teff $\geq$ 20,000 K) DA white dwarf stars for
which either \ghrs, \stis\ or high resolution \iue\ echelle
spectra were available. In each star, the resonance doublets of
\civ, \nv\ and \siiv\ have been examined for signs of multiplicity
(whether in the form of asymmetry or distinct components), and
statistical tests applied to determine the significance of
proposed secondary features. In several stars, Gaussian profiles
are used to model the observed lines. A Gaussian approximation to
line data can be justified since, at the resolution of both the
IUE echelle data (R$\simeq$20,000) and the STIS E140M grating
(R$\simeq$40,000), unsaturated ISM lines are not resolved, nor are
the photospheric profiles of most absorption lines. In some cases,
elemental abundances are considered as additional evidence for the
presence of non-equilibrium processes. The discussion of
individual objects is divided into two subsections, covering stars
which show clear (or suspected) highly ionised, non-photospheric
features, followed by those that do not {\em at the resolution of
the data used in this study}. Several of the stars in this
category exhibit unusual features which require further
investigation at higher spectral resolution.

In section~\ref{sec:discussion}, the results of this study are
discussed, and a variety of explanations for the presence of
highly ionised non-photospheric features in certain white dwarfs,
are considered.

\section{Observations and analysis}\label{sec:obs}
\subsection{Observational data}
The sample of stars was chosen to match that of~\citet{Bar:2000},
with the addition of the super-hot DA PG 0948+534 and Ton 021,
since \ghrs, \stis\ or high resolution \iue\ echelle spectra were
available for these objects. Table~\ref{tab:samplestars}
summarises the survey stars, their basic physical parameters, and
the source(s) of data used in this work. Values for temperature
and gravity were taken from~\citet{Bar:2000}. The adopted visual
magnitudes were those of~\citet{Mar:1997a}, except where otherwise
stated. The mass, radius and distance to each star was estimated
using the evolutionary models developed by~\citet{Woo:1995},
taking the stated values of\linebreak \teff, \lg\ and \mv\ as
input parameters.

\begin{table*}
\begin{minipage}{190mm}\caption[Stars included in the survey of circumstellar
features]{Stars included in the survey of circumstellar features,
in descending temperature order. Mass, radius and distance
calculated from the evolutionary models of~\citet{Woo:1995}.
Visual magnitudes taken from~\citet{Mar:1997a} unless otherwise
stated.}
\begin{center}
\begin{tabular}{l c c c c c c c c} \hline \hline
Star & \teff (K) & \lg & M (\Msun) & R (\Rsun) & m$_v$ & D (pc) &
Data source {\scriptsize [Mode]} & Resolution \\ \hline PG
0948+534 & 90,000 & 7.27 & 0.478 & 0.027 & 13.71$^{e}$ & 193.8 & \stis\ {\scriptsize [E140M]} & 40000 \\ %
REJ 1738+665 & 71,300 & 7.53 & 0.540 & 0.024  &14.61 & 243& \stis\ {\scriptsize [E140M]} & 40000 \\ %
Ton 021 & 69,700 & 7.47 & 0.550 & 0.023 & 14.53$^{g}$ & 217 & \stis\ {\scriptsize [E140M]} & 40000 \\ %
REJ 0558-373 & 63,000 & 7.66 & 0.580 & 0.019 & 14.37 & 295&\stis\ {\scriptsize [E140M]} & 40000 \\ %
REJ 2214-492 & 62,100 & 7.23 & 0.464 & 0.027 & 11.71 & 69&\iue\ {\scriptsize [SWP]} & 20000\\ %
REJ 0623-371 & 59,700 & 7.00 & 0.399 & 0.033 & 12.09 & 97&\iue\ {\scriptsize [SWP]} & 20000\\ %
WD 2218+706 & 56,900 & 7.00 & 0.396 & 0.033 & 15.40$^{c}$ & 436 &\stis\ {\scriptsize [E140M]} & 40000 \\ %
Feige 24 & 56,400 & 7.36 & 0.546 & 0.021 & 12.56 & 78 &\stis\
{\scriptsize [E140M]} & 40000\\ REJ 2334-471 & 54,600 & 7.58 &
0.536 & 0.020 & 13.44 & 104&\iue\ {\scriptsize [SWP]} & 20000\\
G191-B2B & 54,000 & 7.39 & 0.510 & 0.021 & 11.73 & 50&\stis\
{\scriptsize [E140M]} & 40000 \\ GD 246 & 53,700 & 7.74 & 0.646 &
0.016 & 13.09 & 72&\stis\ {\scriptsize [E140M]}/\iue\ {\scriptsize
[SWP]} & 40000/20000\\ REJ 0457-281 & 53,600 & 7.80 & 0.612 &
0.016 & 13.95 & 108&\iue\ {\scriptsize [SWP]} & 20000
\\ PG 1123+189 & 52,700 & 7.52 & 0.511 & 0.021 & 14.13 & 147 &\stis\ {\scriptsize [E140H]} & 110000\\ HZ 43 & 49,000 &
7.90 & 0.561 & 0.018 & 12.99 & 71&\iue\ {\scriptsize [SWP]} &
20000
\\
REJ 1032+532 & 46,300 & 7.78 & 0.585 & 0.016 & 14.46 & 127&\stis\
{\scriptsize [E140M]} & 40000
\\ REJ 2156-546 & 45,900 & 7.74 &0.568 & 0.017 & 14.44 & 129
&\stis\ {\scriptsize [E140M]} & 40000\\ PG 1057+719 & 39,555 &
7.66 & 0.519 & 0.018 & 14.68$^{b}$ & 411& \ghrs\ {\scriptsize
[G160M]} & 22000
\\ REJ 1614-085 & 38,500 & 7.85 & 0.596 & 0.015 & 14.01 &
86&\ghrs\  {\scriptsize [G160M]} & 22000\\ GD 394 & 38,400 & 7.84
& 0.592 & 0.015 & 13.08 & 57&\iue\ {\scriptsize [SWP]}/\ghrs\
{\scriptsize [G160M]} & 20000/22000
\\ GD 153 & 37,900 & 7.70 & 0.520 & 0.017 & 13.35$^{a}$ & 73&\iue\ {\scriptsize [SWP]} & 20000\\ GD 659 & 35,300 & 8.00 & 0.662 & 0.013 &
13.37 & 53&\stis\ {\scriptsize [E140M]}/\iue\ {\scriptsize [SWP]}
& 40000/20000
\\ EG 102 & 20,200 & 7.90 & 0.573 & 0.014 & 12.76$^{d}$ & 25&\iue\ {\scriptsize [SWP]} & 20000 \\ Wolf 1346 & 20,000 & 7.90 & 0.572 & 0.014 &
11.52$^{f}$ & 14&\iue\ {\scriptsize [SWP]} & 20000
\\ \hline
\multicolumn{2}{l}{{\footnotesize $^{a}$~\citet{Boh:1995}}} &
\multicolumn{3}{l}{{\footnotesize $^{b}$~\citet{Hol:1997b}}} &
\multicolumn{3}{l}{{\footnotesize $^{c}$~\citet{Nap:1995}}} \\
\multicolumn{2}{l}{{\footnotesize $^{d}$~\citet{Gre:1984}}} &
\multicolumn{3}{l}{{\footnotesize $^{e}$~\citet{Hol:1995b}}} &
\multicolumn{3}{l}{{\footnotesize $^{f}$~\citet{Dah:1988}}} \\
\multicolumn{2}{l}{{\footnotesize $^{g}$~\citet{Saf:1998}}} \\

\end{tabular}
\label{tab:samplestars}
\end{center}
\end{minipage}
\end{table*}

The results of the survey are summarised in
table~\ref{tab:measured}, which includes the measured velocities
of interstellar (\vism), photospheric (\vphot), and any
non-photospheric highly ionised lines (\vcirc) for all stars in
the sample. Note that values for \vcirc\ are not relative to the
photospheric features, but are absolute velocities.  Also included
in the table are estimated values for the escape velocity (\vesc)
and gravitational redshift (\vgr) of each star, and the velocity
of the primary component of the local interstellar cloud along the
line of sight to each star (\vlic). As noted previously, the
spectral resolution of \iue, \ghrs\ and \stis\ (in the E140M
configuration prevalent in this study) is insufficient to
completely resolve the ISM components, and hence the value of
\vism\ presented in table~\ref{tab:measured} represents the
velocity of the primary component (or blend) observed in the data.
This table also includes calculated values for gravitational
redshift, escape velocity, and the velocity of the principal
component of the local interstellar cloud in that direction. The
local interstellar cloud can be described with reasonable accuracy
by a cloud moving at $26 \pm 1$ \kms\ (heliocentric velocity)
towards \lii\ = ($186 \pm 3$)\deg, \bii\ = ($-16 \pm 3$)\deg, or
\ra~=~74.5\deg, \dec~=~+15\deg~\citep{Lal:1995}. The velocity of
absorption lines from the LIC in a particular star, \vlics, may be
estimated from the projection of \vlic\ onto the target direction.

\begin{table*}
\begin{minipage}{190mm}\caption[Measured velocities for interstellar, photospheric and
non-photospheric lines for surveyed stars]{Measured velocities for
interstellar, photospheric and non-photospheric lines for surveyed
stars. Tentative identifications are\\ marked ``?''. All
velocities in \kms.}
\begin{center}
\begin{tabular}{l r r r r r r} \hline \hline
Star & \vesc &  \vgr  & \vlic & \vism & \vphot & \vcirc \\ \hline
PG 0948+534 & 2645 & 11.3 & 10.00 & -0.26 \err 1.26 & -14.25 \err\ 0.22 \\ %
REJ 1738+665 & 3365 & 14.2 &-3.46 & -18.56 \err 0.09 & 30.19 \err0.27 & -17.80 \err 0.33 \\ %
Ton 021 & 3074 & 15.2 & 10.61 & 0.86 \err 0.01 & 37.14 \err\ 0.21 & 7.62 \err 1.27 \\ %
REJ 0558-373 & 3477 & 19.6 &25.19 &11.61 \err 1.40 & 22.74 \err 2.81 & 7.15 \err\ 0.77\\ %
REJ 2214-492 & 2526 & 10.9 &-8.54 & -1.72 \err0.51 & 33.49 \err0.45 \\ %
REJ 0623-371 & 2143 & 7.7 &14.46 & 16.40 \err0.70 & 41.15 \err0.56 \\ %
WD 2218+706 & 2143 & 7.6 &4.93 & -10.11 \err 0.19 & -38.69 \err 0.23 & -16.34 \err 0.67 \\ %
Feige 24 & 2588 & 17.0 &20.52 & 8.22 \err 0.10 & 80.33 \err 0.50 &$\Delta$V=25.0 \err 3.5 \\ %
REJ 2334-471 & 3253 & 17.0 & -2.62 & 15.15 \err\ 1.52 & 40.57 \err\ 2.07 \\ %
G191-B2B$^{*}$ & 2679 & 15.4 &20.58 & 8.6 & 25.20 \err 0.26 & 7.56 \err 0.19 \\ %
GD 246 & 3498 & 25.7 &2.40 & -5.78 \err0.12 & -13.29 \err0.25 \\ %
REJ 0457-281 & 3749 & 24.3 & 18.92 & 14.48 \err2.15 & 76.91 \err 0.83 & 21.76 \err 1.27 \\ %
PG 1123+189 & 3111 & 15.5 & -0.36 & -0.67 \err0.06 & 12.55 \err0.53 \\ %
HZ 43 & 4461 & 19.8 & -8.83 & -16.86 \err2.07 & ---- \\ %
REJ 1032+532 & 3663 & 23.3 & 7.41 & 0.84 \err0.21 & 38.16 \err0.40 \\ %
REJ 2156-546 & 3606 & 21.2 & -9.65 & -8.39 \err0.18 & -19.94 \err 0.68 & -1.65 \err 0.76 ?\\ %
PG 1057+719 & 3384 & 18.3 & 6.51 & -2.89 \err0.69 & 75.35 \err2.59  \\ %
REJ 1614-085 & 3845 & 25.3 &-25.29 & -29.56 \err0.33 & -37.31 \err 0.40 & -71.42 \err 6.63 \\ %
GD 394 & 3801 & 25.1 & -2.16 & -7.28 \err1.42 & 28.75 \err0.91 \\ %
GD 153 & 3444 & 19.4 & -8.65 & -8.42 \err 2.68 & 12.45 \err2.07 ? \\ %
GD 659 & 4254 & 32.4 & 6.12 & 9.77 \err0.22 & 34.28 \err 0.17 & -2.97 \err 3.00 ? \\ %
EG 102 & 3934 & 26.0 & 1.05 & -1.37 \err 3.85 & -0.69 \err 1.44\\ %
Wolf 1346 & 3934 & 26.0 & -10.76 & -15.38 \err 0.95 & 24.32 \err 1.41 \\ %
\hline
\multicolumn{7}{l}{~$^{*}${\footnotesize ISM velocity from
~\citet{Sah:1999}; no error quoted. Value for ISM velocity
determined from\vspace{-0.75mm}}} \\%
\multicolumn{7}{l}{{\footnotesize~\ \ \vspace{-0.75mm}current work
(using lower resolution data) is $v_{\hbox{\tiny{ISM}}} = 16 \pm
1$ kms$^{-1}$.}}
\end{tabular}
\label{tab:measured}
\end{center}
\end{minipage}
\end{table*}

\subsection{Analysis methods}
Absorption line parameters were measured using an IDL code,
``Lines'', written by one of us (JBH). The routine measures the
properties (wavelength, velocity, equivalent width and associated
uncertainties) of cursor-defined features in an input spectrum,
given a user-supplied rest-frame wavelength for the feature. The
principal lines in each spectrum were identified using the data
contained in HBS, beginning with unambiguous features such as the
saturated interstellar lines of, e.g., \NI, and the resonance
doublets of photospheric \civ\ and \siiv. Subsequent
identifications were validated against the resulting ISM or
photospheric velocity.

Gaussian line profiles have been fitted to observed features to
determine the velocity of circumstellar material, or multiple ISM
clouds along the line-of-sight. The profiles were fitted using
further bespoke IDL routines which applied first a single, and
then dual, Gaussian to cursor-identified features in the spectra
via the $\chi^2$ minimisation technique, generating values for the
velocity and equivalent width of the best-fit components. In cases
where the dual-Gaussian fit was not obviously superior, $\chi^2$
values from the single and dual fits were then used to perform a
standard {\sl F}-test, following the method outlined
by~\citet{Hol:1997b}, to determine whether a significantly better
fit to observation was obtained with the latter.

In addition to the IDL routines, cross-checking of results was
performed using the {\em Dipso} package (part of a suite of tools
produced for the UK {\em Starlink} system). Excellent consistency
was observed between results obtained from these disparate
packages.

Co-addition of spectral features has been performed in several
cases in order to reveal details which are only marginally
detectable at the S/N of the data. In this technique, $\sim$ 10
\AA\ - wide sections of spectral data are extracted, each centered
on the wavelength of a particular line of a given species (e.g.
the \lam\lam 1548,1550 lines of \civ). The sections are then
transformed into velocity space so that each shows a line at the
velocity of the primary interstellar cloud or the photosphere.
Spectral sections are then co-added so that the strength of the
(randomly distributed) noise features remains essentially
unchanged, while absorption features sharing a common velocity are
summed. The result is a spectrum with improved S/N, showing the
profile of lines of a particular species.

Co-addition does not improve the resolution of the data, and is
therefore ineffective in revealing circumstellar features which
are blended with their photospheric counterparts at the resolution
of the data. Further, the technique is only effective when several
of the primary lines are accompanied by such features.
Nevertheless, co-addition has proved to be a useful technique in
detecting weak circumstellar features which are clearly separated
from the primary component at the resolution of the instrument.

\subsection{Comments on individual objects}
\subsubsection{Stars exhibiting circumstellar features}

{\em REJ 1738+665}\\ \label{sec:rej1738} REJ 1738+665 is the
hottest DA white dwarf to be detected by {\em
ROSAT}~\citep{Bar:1994}. A photospheric velocity of \vphot\
$\approx\ 30 \pm 1$ \kms\ is determined, based on absorption
features arising from Fe, Ni and O which show no multiple
components; interstellar lines indicate a line of sight ISM
velocity of \vism\ $\approx\ -18 \pm 1$ \kms. The line of sight
velocity of the LIC is estimated to be \vlics\ $\approx\ -3.4$
\kms.

\begin{figure}
\centering
\rotatebox{270}{\resizebox{!}{8.5cm}{\includegraphics{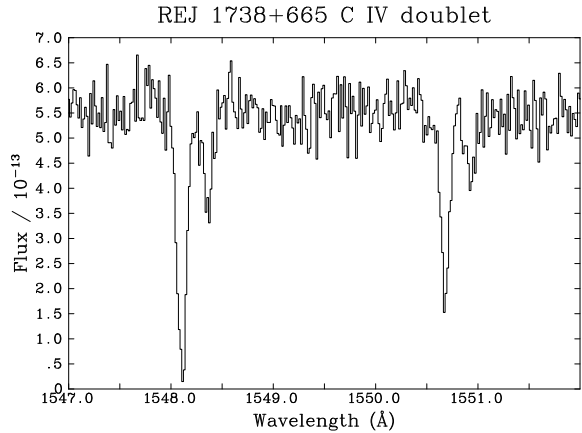}}
} \vspace{0mm}\caption{Clear multiplicity in the C {\scriptsize
IV} doublet of RE 1738+665, with the photospheric components
dominated by the blueshifted lines.} \label{fig:rej1738}
\end{figure}

\begin{figure}
\begin{minipage}{8cm}
\resizebox{1.05\textwidth}{!}
{\rotatebox{0}{\includegraphics{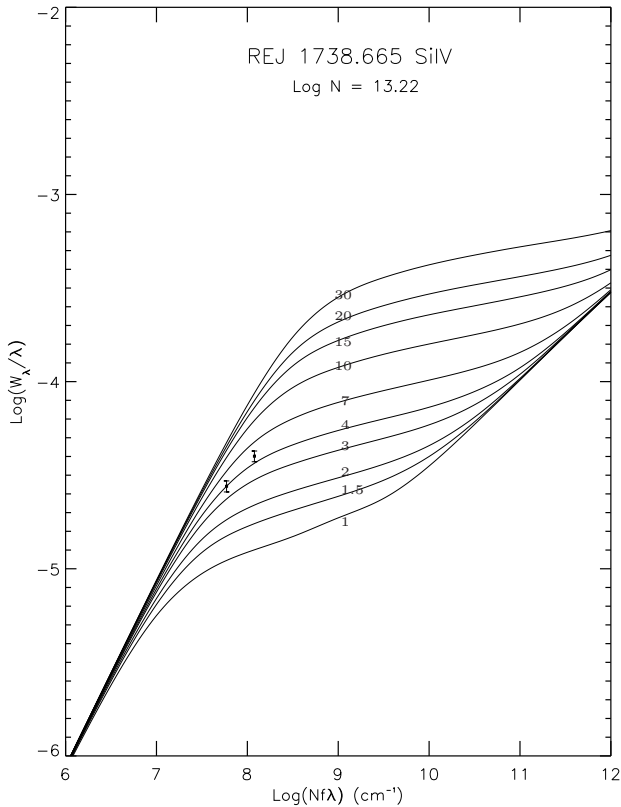}}}
\put(-112,189){\tiny{30}}%
\put(-112,181){\tiny{20}}%
\put(-112,171){\tiny{15}}%
\put(-112,162){\tiny{10}}%
\put(-112,149){\tiny{\ 7}}%
\put(-112,140){\tiny{\ 4}}%
\put(-112,132){\tiny{\ 3}}%
\put(-112,122){\tiny{\ 2}}%
\put(-112,115){\tiny{\ 1.5}}%
\put(-112,103){\tiny{\ 1}}%
\end{minipage}
\caption{Curves of growth for \siiv\  in REJ 1738+665. Separate
curves in each plot correspond to different values of the Doppler
parameter, $b$ (indicated in \kms).} \label{fig:rej1738_cog}
\end{figure}

Clear evidence is found for the presence of circumstellar material
in this star. Figure~\ref{fig:rej1738} shows the \civ\ resonance
doublet in REJ~1738+665, with shifted components at $-18.5 \pm
0.5$ \kms\ dominating the 30 \kms\ photospheric contribution.
Shifted features with similar velocities are observed in several
other species, although in each case the photospheric component is
dominant. The \siiv\ doublet shows non-photospheric components at
$-17.7 \pm 0.7$ \kms. Viewed individually, the lines of the \nv\
doublet ($\lambda\lambda$ 1238.821, 1242.804) show no evidence of
companions, but co-addition of these features suggests an extra,
weak component at $-15.2 \pm 2$ \kms. The \oiv\ doublet
($\lambda\lambda$1338.612,1343.512) shows no additional features,
but the \ov\ line at $\lambda$1371.292 is accompanied by a weak
shifted component at $-18.7 \pm 0.5$ \kms. Curve-of-growth
analysis of the \civ\ and \siiv\ lines indicate column densities
of N (\civ) = $1.70 \times 10^{13}$ atoms cm$^{-2}$ and N (\siiv)
= $1.66 \times 10^{13}$ atoms cm$^{-2}$ with $b \approx 5$ \kms\
(curves for the \siiv\ features are shown in
figure~\ref{fig:rej1738_cog}).

The velocities measured for these shifted features differ from
\vism\ by less than 0.75 \kms. This raises the possibility that
they may be produced by photoionisation of the ISM within the
Str\"omgren sphere around the star. However,~\citet{Twe:1994} also
present evidence for the possible existence of a planetary nebula
around the star, based on the observation of \NII\ circumstellar
features at optical wavelengths. The non-photospheric features of
REJ~1738+665 may therefore be produced by ionisation of the
ancient planetary nebula remnant surrounding this star. The
relationship between planetary nebul\ae\ and highly ionised
non-photospheric lines is discussed in
section~\ref{sec:discussion}.\\

{\em Ton 021}\\ Other than its inclusion in general white dwarf
surveys, Ton 021 has received comparatively little attention in
the literature. We determine \teff = 69700 \err\ 530 K and \lg =
7.47 \err\ 0.05 (c.f. 69711 \err\ 1030 and 7.469 \err\ 0.05
respectively from ~\citet{Fin:1997}). Weighted averages of the
significant interstellar and photospheric absorption features give
values of \vism\ = 0.86 \err\ 0.01 \kms\ and \vphot\ = 37.25 \err\
0.22 \kms. Among the observed interstellar features are strong
lines of \cii\ at $\lambda$1334.5323 (equivalent width 171 \err\
1.6 m\AA) and $\lambda$1335.7076 (equivalent width 61 \err\ 2.2
m\AA). Velocities consistent with the weighted average \vism\ are
observed for the broad interstellar lines of \NI\ and \siii;
however, the \siiii\ $\lambda$1206.5 line appears at 7.8 \err\ 0.7
\kms. This is not a unique observation;~\citet{Hol:1999b} find
significant differences between the velocity of this line and the
mean value of \vism\ in REJ 1032+532 (also included in the current
study).~\citeauthor{Hol:1999b} discuss this observation in some
detail, arguing that the \siiii\ feature is unlikely to originate
in the LIC due to the high ionisation fraction of hydrogen ($\sim
95$\%) required to maintain detectable amounts of \siiii, which
has a high rate coefficient for charge exchange with neutral H. As
noted by~\citet{Hol:1999b}, a similar observation is made
by~\citet{Vid:1998} in the case of G191-B2B.

\begin{figure}
\centering
\rotatebox{270}{\resizebox{!}{8.5cm}{\includegraphics{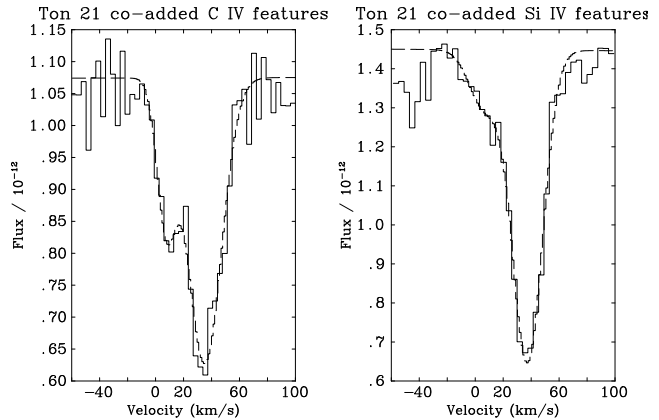}}
} \caption{The co-added C {\scriptsize IV} and Si {\scriptsize IV}
resonance lines of Ton 021. Clear evidence for multiplicity is
observed in the C {\scriptsize IV} features, while the presence of
a non-photospheric component to the Si {\scriptsize IV} features
is less compelling. Dotted line: best fit dual-Gaussian to the
data.} \label{fig:ton21_coadded}
\end{figure}

Both lines of the \civ\ resonance doublet are accompanied by
shifted features. In the $\lambda$1548.202 line, the
non-photospheric component is best fit by a Gaussian with a
velocity of \vcirc\ = 9.5 \err\ 1.0 \kms, and an equivalent width
of 20 \err\ 5 m\AA. This velocity agrees, within the stated error
margin, with that determined for the \siiii\ line. The shifted
component in the $\lambda$1550.774 line is fit by a Gaussian at
\vcirc\ = 5.5 \err\ 2.2 \kms, with an equivalent width of 9.4
\err\ 2.5 m\AA. The estimated \civ\ column density contributing to
these shifted features is N (\civ) = $3.98 \times 10^{12}$ --
$1.26 \times 10^{13}$ atoms cm$^{-2}$ based on a curve-of-growth
analysis. In both cases, the photospheric components are found at
velocities consistent with the average value for \vphot. The
presence of a non-photospheric \civ\ feature is most clearly
demonstrated in the co-added lines of the resonance doublet, as
shown in figure~\ref{fig:ton21_coadded}.

Viewed individually, the lines of the \siiv\ resonance doublet
appear slightly asymmetrical but show no clear multiplicity at the
resolution of this data. When co-added, this asymmetry is more
noticeable (as shown again in fig~\ref{fig:ton21_coadded}), and we
find the best fit dual-gaussian to be one with a primary component
at $\sim 37 \pm\ 1.2$ \kms, in agreement with \vphot, and a
non-photospheric component at $\sim 6.5 \pm\ 12.3$ \kms. Although
this velocity is close to the value of \vcirc\ found in the \civ\
doublet, there is considerable uncertainty in the measurement, and
data of higher S/N will be required to confirm the detection of
non-photospheric components to the \siiv\ doublet of Ton 021.\\

{\em REJ 0558-373}\\ The \civ\ $\lambda\lambda$1548.202,1550.774
lines in this star are moderately asymmetrical. An {\sl F}-test
suggests that a dual Gaussian fit is preferred over a single line,
above the 98\%\ confidence interval, for each feature. No
corresponding asymmetries are observed elsewhere in REJ~0558-373.

\begin{figure}
\centering
\rotatebox{270}{\resizebox{!}{9cm}{\includegraphics{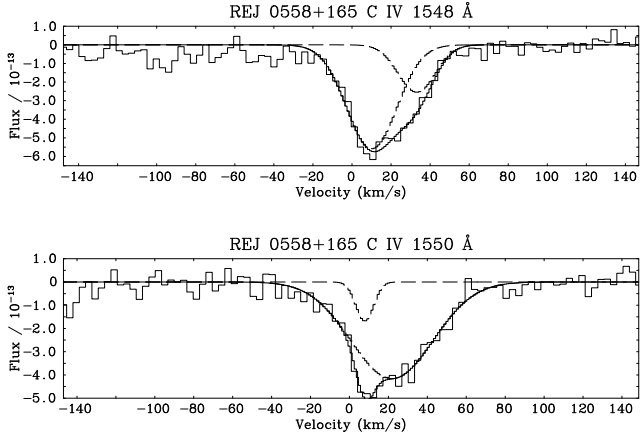}}
} \vspace{-0mm}\caption{Lines of the C {\scriptsize IV} doublet of
REJ 0558-373, in velocity space, with Gaussian fits overlaid
(histogram = \stis\ data; dashed curves: component Gaussians;
solid curve: compound fit).} \label{fig:rej0558}
\end{figure}

For the $\lambda$1548.202 line, the individual components are
found at $7.0 \pm 1.0$ and $26.8 \pm 1.2$ \kms\ (with equivalent
widths of 79 and 68 m\AA\ respectively). Corresponding values for
the $\lambda$1550.774 line are $7.9 \pm 1.2$ and $21.8 \pm 1.0$
\kms\ (13 m\AA\ and 150 m\AA). Fits to each line are illustrated
in figure~\ref{fig:rej0558}. For comparison, \vphot\ = 22.7 \err\
2.8 \kms, and \vism\ = 11.6 \err\ 1.4 \kms, and hence the longer
wavelength component of each \civ\ line is in reasonable agreement
with the photospheric value.

\vgr\ is estimated as $\approx$ 20 \kms\ for this star, which is
greater than the $\sim$ 10 \kms\ difference between the
photospheric and shifted \civ\ components. It is therefore
possible that the blueshifted, non-photospheric \civ\ features
could be formed by material {\em within} the potential well,
rather than the weakly shifted outer regions. However, the
non-photospheric components lie close to the velocity of the ISM,
raising instead the possibility that the star is ionising material
in its local interstellar environment. In either case, the absence
of corresponding features in other strong lines is puzzling.\\

{\em WD 2218+706}\\ This object is unusual and important in
several respects. Lines of the \civ\ and \siiv\ doublets are
clearly multiple, dominated by a photospheric contribution, but
with accompanying components of comparable equivalent width at a
velocity of -16.3~\err\ 0.7 \kms. These features are therefore
{\em redshifted} with respect to the photospheric velocity
(\vphot~=~-38.7~\err~0.2~\kms), possibly representing the infall
of material onto the white dwarf. Gravitational redshift therefore
provides no viable explanation for these lines.

\begin{figure}
\centering
\rotatebox{270}{\resizebox{!}{8.5cm}{\includegraphics{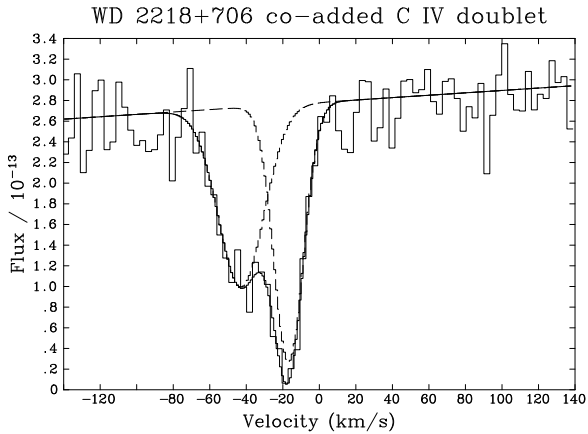}}
} \caption{Histogram: co-added lines of the \civ\ doublet in WD
2218+706 showing the photospheric component at 38.8 \kms,
accompanied by a redshifted feature at -16.3 \kms. The data are
accompanied by the individual and summed components of the
best-fit dual Gaussian.} \label{fig:2218_civ}
\end{figure}

WD 2218+706 is surrounded by an old planetary nebula, DeHt5, and
is discussed by~\citet{Nap:1995}. In a study of planetary nebula
dynamics,~\citet{Dga:1998} show that in regions where the ISM is
reasonably dense (such as the galactic plane), Rayleigh-Taylor
instabilities can develop in the outer regions of planetary
nebul\ae, leading to fragmentation of the halo, and allowing the
surrounding ISM to pass into the inner regions of the nebula where
photoionisation can occur. Although WD 2218+706 is out of the
galactic plane ($b_{\hbox{\scriptsize{II}}} = 11.6$\deg), and
therefore lies in a region where the mean ISM density may be
expected to be relatively low,~\citet{Kun:1998} describes the
morphology of a nearby giant molecular cloud complex consisting of
a large number of distinct regions previously identified in
independent surveys; several are found close to WD~2218+706, and
two are of particular interest (Lynds 1217 and Lynds 1219). The
central portions of these clouds have galactic coordinates within
less than 0.5\deg\ of this star, and their distance limits (from
380 to 450 pc) encompass the distance to WD 2218+706 (440 pc,
from~\citeauthor{Nap:1995}). This raises the possibility that the
star may lie in an area where the ISM is particularly dense,
allowing instability and inflow to take place (a curve of growth
analysis for the non-photospheric features in WD~2218+706 suggests
column densities of N (\civ) = $4.17 \times 10^{13}$ atoms
cm$^{-2}$ and N (\siiv) = $4.07 \times 10^{13}$ atoms cm$^{-2}$,
each with a Doppler parameter of 6 \kms). However, alternative
explanations, such as the presence of a hidden companion, are also
deserving of investigation, and this work is currently in
progress.

\begin{figure}
\begin{minipage}{8cm}
\resizebox{1.05\textwidth}{!}{\rotatebox{0}{\includegraphics{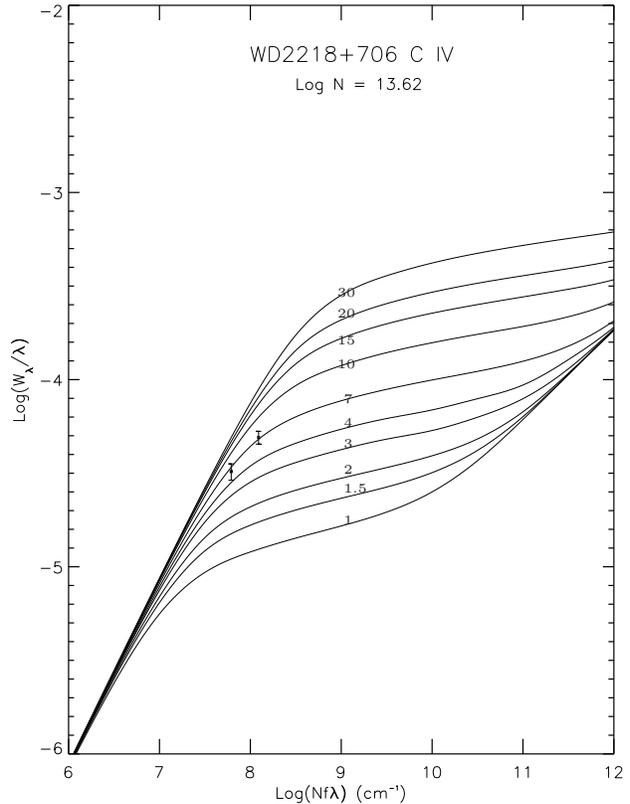}}}
\put(-112,189){\tiny{30}}%
\put(-112,181){\tiny{20}}%
\put(-112,171){\tiny{15}}%
\put(-112,162){\tiny{10}}%
\put(-112,149){\tiny{\ 7}}%
\put(-112,140){\tiny{\ 4}}%
\put(-112,132){\tiny{\ 3}}%
\put(-112,122){\tiny{\ 2}}%
\put(-112,115){\tiny{\ 1.5}}%
\put(-112,103){\tiny{\ 1}}%
\end{minipage}
\caption[Curves of growth for the redshifted \civ\ features in WD
2218+706]{Curves of growth for the redshifted C {\scriptsize IV}
features in WD 2218+706. Separate curves in each plot correspond
to different values of the Doppler parameter, $b$ (indicated in
\kms)} \label{fig:wd2218_cog}
\end{figure}

During the course of the WD 2218+706 study, evidence was found for
the existence of trace amounts of He in the \stis\
spectrum~\citep{Bar:2001}, with the \heii\ $\lambda$1640.5050 line
observed in the \stis\ spectrum close to the estimated
photospheric velocity. As noted by~\citeauthor{Bar:2001}, the
$\lambda$1640.5050 line has an n=2 lower level, which should not
be populated in collisionless interstellar material, while any
helium in the surrounding planetary nebula would be expected to be
found in emission. Hence a photospheric origin appears to be the
most satisfactory of these three possible sources. The surface
gravity of WD 2218+706 (\lg $\approx$ 7.00) is low for an isolated
white dwarf, and may be explained in terms of close-binary
evolution, in which the progenitor star fills its Roche lobe and
loses mass to a companion. This loss of material prevents helium
ignition from taking place. Instead, the star, consisting of a He
core surrounded by a H-rich envelope (which still supports nuclear
reactions at the base), contracts slowly towards the low-mass,
He-core white dwarf configuration~\citep{Dri:1998,Nap:1999}. As
the case of Feige 24 illustrates, the presence of a binary
companion can contribute to the appearance of circumstellar lines.
However, there is no direct evidence for the existence of a binary
companion, and the range of possible masses for this star does not
preclude AGB evolution, leaving open the possibility that WD
2218+706 is the product of single-star evolution.

Evidence exists for at least one other H-rich white dwarf star
exhibiting the \heii\ \lam 1640 line, in the DAB HS 0209+0832
(\teff $\sim$ 35,000 K; \lg $\sim $ 7.8). This star is discussed
by~\citet{Wol:2000}, who suggest that significant quantities of He
are present in the atmosphere, despite the short diffusion
timescales for He, as a result of ongoing accretion of matter from
an interstellar cloud. Supporting evidence for this explanation
can be found in the work by~\citet{Heb:1997}, who observe
variability in the strength of the He \lam 1640 line, possibly as
a result of the passage of HS 0209+0832 through an inhomogeneous
medium. Alternatively,~\citet{Ung:2000} find that DAO stars, in
which mass loss in the form of a stellar wind prevents He from
sinking out of the atmosphere, can transform into DA stars when
the phase of wind-driven mass loss ends. This transition is found
to occur near \lg $\sim$ 7.0 for a star with \teff $\sim$ 60,000 K
(see fig.6, ~\citet{Ung:2000}), raising the possibility that WD
2218+706 may be such a transitional object.\\

{\em Feige 24}\\ Feige 24 is a white dwarf+red dwarf binary
system, and is the subject of several detailed studies
(e.g.~\citealp{Dup:1982,Ven:1994}). Two \stis\ data-sets were
available for this star, acquired on November
29$^{\hbox{\small{th}}}$ 1997 (binary phase 0.73--0.75) and
January 4$^{\hbox{\small{th}}}$ 1998 (binary phase 0.23--0.25),
representing the orbital quadrature points. \citeauthor{Ven:1994}
estimate a systemic velocity of 62.0 \err\ 1.4 \kms. For the
current study, the systemic velocity has been estimated by taking
the mean of the photospheric values obtained from each data set
(31.6 and 129.1 \kms), resulting in an estimated systemic velocity
of 80.3 \err\ 0.5 \kms. \vism\ is estimated at 8.2 \err 0.1 \kms.

Feige 24 is known to exhibit multiple components, in the lines of
the \civ\ doublet only. In this work, the dominant components
match the photospheric velocity of each data-set, and secondary
features are observed to remain at 7.8 \err\ 0.2 \kms,
irrespective of the orbital phase. This stationary component has
been discussed by~\citeauthor{Dup:1982}, who suggest that the most
likely source is a Str\"omgren sphere excited by the white dwarf,
and measure column densities of  N(\civ) = $3.86 (\pm 1.51) \times
10^{13}$ atoms cm$^{-2}$ using
curves-of-growth.~\citeauthor{Ven:1994} investigated the
possibility that the material responsible resides in the
photosphere, in a circumstellar shell, or in a wind from the red
dwarf companion. The current study shows no shifted components in
any of the other resonance lines (e.g. \siiv\ or \nv). A {\em
third}, very weak feature on the red side of each photospheric
line in the \civ\ doublet, most obviously at 1550 \AA, is
identified as the \fev\ \lam\ 1550.907 line.

\begin{figure}
\centering
\rotatebox{270}{\resizebox{!}{8.5cm}{\includegraphics{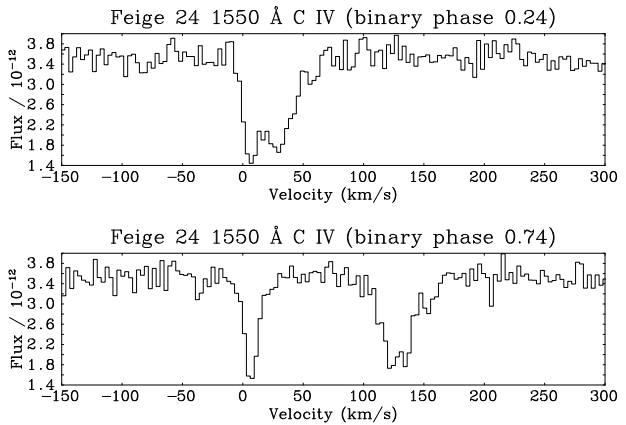}}
} \caption{The C {\scriptsize IV} 1550 \AA\ feature of Feige 24 as
observed at the two quadrature points. The stationary
circumstellar component is clearly visible, as is the orbital
velocity of the photospheric component.} \label{fig:feige24}
\end{figure}

The gravitational redshift estimate made during this study, \vgr\
$\approx 17$ \kms, is somewhat higher than that derived
by~\citeauthor{Ven:1994} (\vgr\ $\approx 9 \pm 2$ \kms), but is
still too low to explain the secondary \civ\ components. However,
their velocities agree, within error, with that of the ISM, and
hence a link between the star and its immediate surroundings
(beyond any circumstellar shell) cannot be discounted.
Alternatively,~\citeauthor{Ven:1994} estimate that a \civ\ column
density from N (\civ) = $7.94 \times 10^{11}$ - $7.94 \times
10^{13}$ atoms cm$^{-2}$ (corresponding to equivalent widths of
between 4--400 m\AA\ for the non-photospheric component of the
\lam\ 1550.774 line, assuming a linear curve-of-growth), would be
consistent with mass loss from the red dwarf companion. Although
insufficient data are available to derive an unambiguous \civ\
column density in this study, the estimated value of N (\civ) =
$1.48 \times 10^{13}$ atoms cm$^{-2}$, is within the range of
possible values obtained by~\citeauthor{Ven:1994}, and the
equivalent width of the \lam\ 1550.774 line (24 m\AA) is also
consistent with the large range allowed by earlier estimates.\\

{\em G191-B2B}\\
\begin{figure*}\begin{minipage}{180mm}
\centering
\rotatebox{270}{\resizebox{!}{5.5in}{\includegraphics{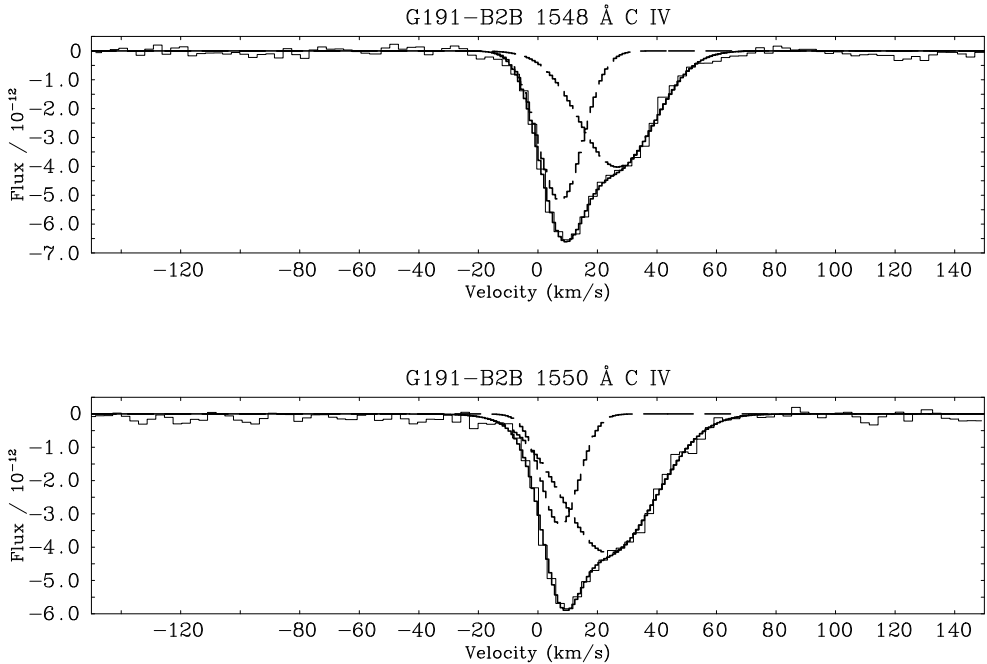}}
} \caption{The C {\scriptsize IV} doublet of G191-B2B, in velocity
space, with compound Gaussian fits (solid lines), and the
individual Gaussian components (dashed lines).} \label{fig:g191}
\end{minipage}\end{figure*}
The \stis\ E140M data for this star show the \civ\ resonance
doublets are accompanied by strong non-photospheric blue shifted
components, confirming the results of~\citet{Bru:1999}. The
wavelength and equivalent width of these features have been
estimated by fitting multiple Gaussians to the data, as shown in
figure~\ref{fig:g191}. The component at photospheric velocity is
of greater equivalent width, and the shifted elements are observed
at velocities of 7.6 \err 0.2 \kms. Recently,~\citet{Hol:2002}
also detected weak blue shifted components to the \siiv\ \lam\lam
1393.755,1402.777 doublet at the same velocity as the
non-photospheric \civ\ features, using 22 co-added \stis\ E140H
spectra.

The calculated figure of \vgr\ $\approx 15.4$ \kms\ is comparable
with the velocity difference between photospheric and shifted high
ionisation features, suggesting that the non-photospheric material
probably resides outside the limit of the potential well. The
velocity of the highly ionised non-photospheric features is
substantially different to the value of the interstellar features
(\vism\ $= 16 \pm 1$ \kms\ as determined in this work), and at
first sight, photoionisation of the cloud responsible for the
primary ISM features does not appear to provide a viable
explanation. However, the value quoted for \vism\ is based on
analysis of an E140M (medium resolution) \stis\ data-set, with a
resolving power of $\sim$ 35000. \citet{Sah:1999} describe
observations made with the E140H grating (resolving power $\sim
110000$), and clearly show {\em two} distinct interstellar
components, with velocities of $\sim$ 8.6 \kms\ and $\sim$ 19.3
\kms, the latter component having a velocity close to the
predicted value of \vlic\ (estimated at 20.58 \kms\ in this
study). Clearly, the highly ionised non-photospheric components
have a velocity which is very close to the 8.6 \kms\ interstellar
cloud.

A curve-of-growth analysis was performed for the \civ\ features in
G191-B2B. As in the case of Feige 24, the availability of only two
datum points prevents any rigorous constraints from being placed
on the implied \civ\ column density, though the value of N (\civ)
= $2.40 \times 10^{13}$ atoms cm$^{-2}$ is not dissimilar from the
results of~\citet{Ven:2001}, who use synthetic modeling techniques
to estimate a value of N (\civ) = $6.31 \times 10^{13}$ atoms
cm$^{-2}$. The Doppler parameter suggested by the current
analysis, $b = 10$ \kms, is significantly higher than the value of
$b = 5.2$ \kms\ presented by~\citeauthor{Ven:2001}; this
discrepancy may also be explained by the lack of available data in
the current work.\\

\begin{figure}
\centering
\rotatebox{0}{\resizebox{8cm}{!}{\includegraphics{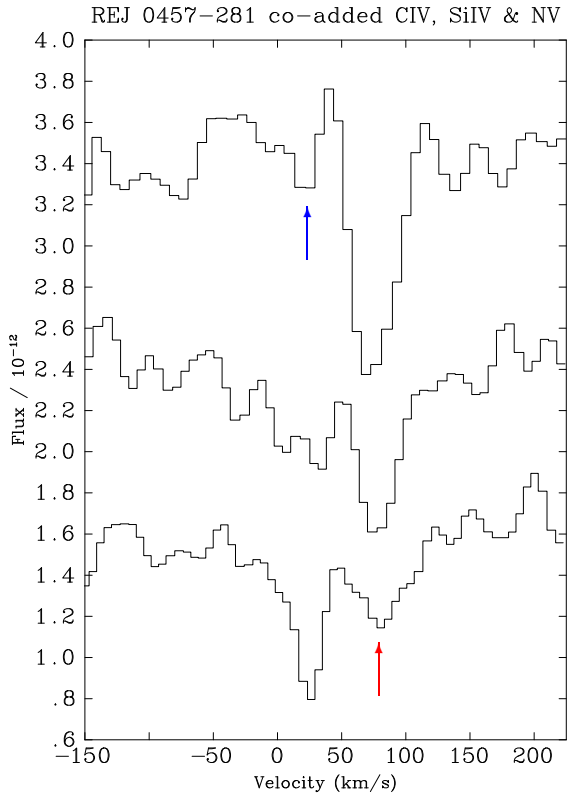}}
} \color{blue} \put(-115,215){\vector(0,2){20}} \color{red}
\put(-88,50){\vector(0,2){20}} \color{black} \caption{From top to
bottom, the co-added lines of the N {\scriptsize V}, Si
{\scriptsize IV} and C {\scriptsize IV} doublets in REJ 0457-281.
Lower arrow indicates the approximate photospheric velocity, upper
arrow shows the position of the shifted components, present in
each species.} \label{fig:rej0457}
\end{figure}

{\em REJ 0457-281}\\ The exceptionally low \hi\ column density to
this star ($1.3 \times 10^{17}$ atom cm$^{-2}$), was revealed in
the discovery paper by~\citet{Bar:1994b}. Along with G191-B2B,
this white dwarf was the first to have phosphorous and sulphur
identified in its spectrum~\citep{Ven:1996}. Later, HBS showed
that the photospheric \siiv\ and \civ\ resonance lines of REJ
0457-281 are accompanied by blue-shifted features
(figure~\ref{fig:rej0457}).

Few interstellar and photospheric lines are identifiable, making
precise velocity measurements difficult. The ISM velocity estimate
of HBS is confirmed, but somewhat higher photospheric velocities
are derived. Co-addition of the \civ\ doublet lines clearly
reveals two velocity components: one at 22.5 \err 1.39 \kms, and
the photospheric component at 81.26 \err\ 2.65 \kms. A curve of
growth analysis for the non-photospheric components suggests N
(\civ) $\sim 1.82 \times 10^{14}$ atoms cm$^{-2}$ with a Doppler
parameter $b \approx 4$ \kms. For the co-added \siiv\ doublet,
corresponding velocities are 19.08 \err\ 4.31 \kms\ and 80.65
\err\ 1.38 \kms. Although multiple velocity components are not
obvious in the \nv\ doublet, there is, nevertheless, evidence to
suggest that they are present (figure~\ref{fig:rej0457}). The
\lam\ 1238.8210 \nv\ line is a narrow, well defined feature at
76.33 \err\ 4.78 \kms, accompanied by a weaker blueshifted feature
at 16.55 \err\ 4.61 \kms. The main \lam\ 1242.804 line has a
similar velocity (76.91 \err\ 4.38 \kms), but shows only tentative
evidence for a blueshifted component.

From these data, a weighted average is computed for the
photospheric and blueshifted velocity components, suggesting
\vphot\ = 76.91 \err\ 0.83, (c.f. 69.60 \err\ 1.97 from HBS), and
\vcirc\ = 21.76 \err\ 1.27. Thus, the estimated velocity shift of
the blueshifted features relative to the photospheric components
agrees, within the stated error, with the 53 \kms\ value of HBS.\\

\begin{figure}
\centering
\rotatebox{270}{\resizebox{!}{8.5cm}{\includegraphics{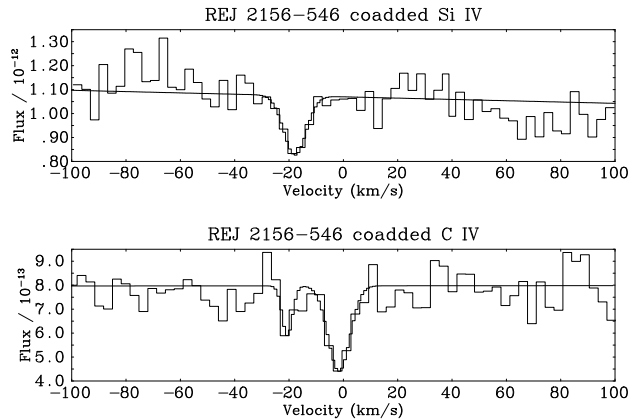}}}
\caption{Co-added Si {\scriptsize IV} and C {\scriptsize IV}
features in REJ 2156-546, with Gaussian fits overlaid.}
\label{fig:rej2156}
\end{figure}

{\em REJ 2156-546}\\ ~\citet{Bar:1997a} determined limits to the
heavy element abundance in REJ 2156-546, and described this object
as being similar to HZ 43 in having a reasonably pure H
atmosphere. The \stis\ spectrum of REJ 2156-546 shows clear
interstellar lines, indicating \vism\ = -8.39 \err 0.17 \kms.
These new, high resolution data also appear to show features due
to photospheric material. The lines are weak, and unambiguous
identifications are limited to the strong resonance doublets of
\siiv\ and \civ, although there may be features from \nv\ and
\niv\ at the detection limit. To obtain a reliable value for
\vphot, the \siiv\ lines (\lam\lam 1393.755,1402.777) were
co-added in velocity space, producing a clear feature at -17.79
\err 1.33 \kms, with an equivalent width of 11 m\AA. Co-addition
of the \civ\ doublet appears to reveal {\em two} features
(figure~\ref{fig:rej2156}). The first, at -20.71 \err 0.80 \kms,
is weak (4.7 m\AA), and very close to our value for \vphot. The
dominant second component lies at -1.65 \err 0.76 \kms, and has an
equivalent width of 20 m\AA. The weighted average of putative
photospheric features produces a value of \vphot\ = -19.94 \err
0.68 \kms.

The proposed \civ\ feature lying at a similar velocity to the
single \siiv\ line is admittedly weak, and spectra of improved
\sn\ will be required before these results can be regarded as
incontrovertible. Nevertheless, if it is assumed that the object
is devoid of any non-photospheric features, whether blue- or
redshifted, then the relatively large difference in velocity
between the \siiv\ line and the dominant \civ\ feature
(approximately 16 \kms) is somewhat difficult to explain. Clearly,
this is an object deserving of further attention.\\

\begin{figure}
\centering
\rotatebox{270}{\resizebox{!}{8.5cm}{\includegraphics{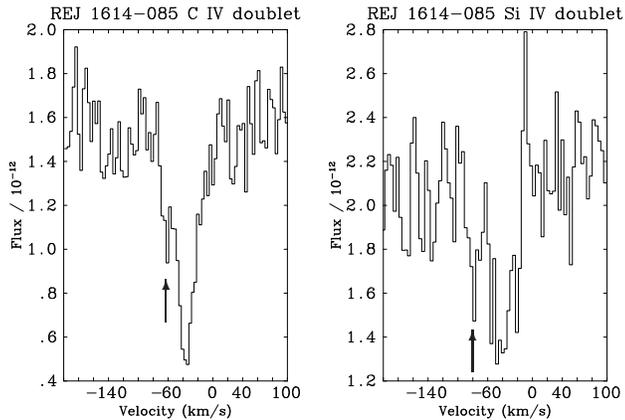}}}
\put(-179,-127){\vector(0,2){16}} \put(-63,-146){\vector(0,2){16}}
\caption[Co-added lines of the \civ\ and \siiv\ doublets in REJ
1614-085, in velocity space]{Co-added lines of the C {\scriptsize
IV} (left) and Si {\scriptsize IV} (right) doublets in REJ
1614-085, in velocity space. The blueshifted components previously
noted by~\citet{Hol:1997b} are indicated by arrows.}
\label{fig:rej1614}
\end{figure}

{\em REJ 1614-085}\\
 \citet{Hol:1997b} found the amount of Si in
the spectrum of REJ 1614-085 to be an order of magnitude
under-abundant compared to the predictions of radiative levitation
calculations, while N appears to be three orders of magnitude
over-abundant. Two velocity components were observed in the line
of sight ISM, but most significant for the current work is the
result that the lines of the \civ\ and \siiv\ doublets exhibit
weak blueshifted features, as illustrated in
figure~\ref{fig:rej1614}.

Results from this study suggest that the primary ISM component
lies at a velocity of -29.56 \err\ 0.33 \kms, and the secondary
component at +48.64 \err\ 1.15 \kms. These values compare
reasonably well with those of~\citeauthor{Hol:1997b}, who find
velocities of -27.05 \err\ 1.5, and +47.40 \err 1.50 \kms\
respectively. Similar agreement with the earlier study is also
found when considering the photospheric and blueshifted features.
The photospheric velocity is found to be \vphot\ = -37.31 \err
0.40 \kms, in excellent agreement with ~\citeauthor{Hol:1997b}. As
recorded previously, no clear evidence exists for a secondary
component to the photospheric \nv\ lines, although it should be
noted that the centroid positions of Gaussian fits to these lines
are found to differ by approximately 5 \kms. In the case of the
\siiv\ doublet, the secondary components are shifted by -40 \kms\
relative to the primary features, as determined
by~\citet{Hol:1997b}; for the \civ\ doublet, this figure is -29
\kms, compared to the value of -25 \kms\ quoted
by~\citet{Hol:1997b}. This apparent discrepancy is most likely to
be a result of the different positions chosen for line demarcation
in the two studies. A curve of growth analysis performed on the
shifted \civ\ features suggests N (\civ) $\approx 3.16 \times
10^{13}$ atoms cm$^{-2}$, and the Doppler parameter $b \approx 2$
\kms, though with only two data points, these values are
particularly poorly constrained, and must not be over
interpreted.\\

{\em GD 659}\\ Both \iue\ and \stis\ data are available for this
star, although the \stis\ spectrum is of limited coverage (1160 --
1357 \AA). The ISM velocity determined from the \iue\ data (\vism\
= 12.33 \err\ 1.52 \kms) agrees with that of HBS, while the
\iue-based photospheric velocity appears to be somewhat lower than
previously quoted (\vphot\ = 33.51 \err\ 1.03 \kms, c.f. 40.31
\err\ 1.83 \kms\ from HBS). \footnote{These features were
originally attributed to circumstellar material, and are listed
accordingly by HBS, based on the 76 \kms\ velocity difference
between the UV features and the -37 \kms\ Balmer line radial
velocity of~\citet{Weg:1974}. The {\em blueshifted}
~\citeauthor{Weg:1974} value is most likely the result of a
typographical error.}

Since these measurements were made with identical data sets, this
discrepancy must be ascribed to differences in choices of
continuum levels and line boundaries used during the measurement
process. However, the available \stis\ data also point to a lower
photospheric velocity, with \vphot\ = 34.28 \err 0.17 \kms, in
agreement with the \iue\ estimate made in this study, and close to
the value of 33.58 \kms\ determined by ~\citet{Hol:2000} using
\stis\ data. The \stis\ ~ISM velocity is also lower than the \iue\
value, at \vism\ = 9.77 \err\ 0.22 \kms. These \stis\ ~velocities,
based on higher resolution data with better \sn, are adopted for
GD 659 in table~\ref{tab:measured}. Although the resonance
doublets of \nv, \siiv\ and \civ\ are clearly visible in the \iue\
data, the resolution is insufficient to observe well defined
Gaussian profiles, and thus the sensitivity to any
non-photospheric components is low. However, the profiles appear
to be narrow, ruling out any obvious multiplicity in these
lines.~\stis\ data show the lines of the \nv\ doublet as narrow
and symmetrical, effectively ruling out the existence of
non-photospheric \nv\ components.

\begin{figure}
\centering

\rotatebox{270}{\resizebox{!}{8.5cm}{\includegraphics{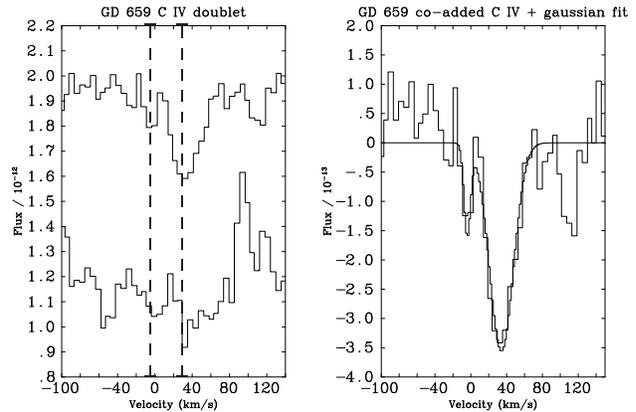}}
} \put(-184,-149){\dashbox{4}(0.0,133)}
\put(-172,-149){\dashbox{4}(0.05,133)}

\color{black}\caption{Left: the \lam\ 1548.202 (top) and \lam\
1550.774 (bottom) C {\scriptsize IV} features in GD 659. Dashed
vertical lines indicate the position of a possible secondary
component near 0 \kms, and the primary photospheric line. Flux
levels have been adjusted by a constant to allow plotting of
features in the same panel. Right: co-addition of the C
{\scriptsize IV} features, with a best-fit double Gaussian profile
overlaid.}\label{fig:gd659}
\end{figure}

One interesting feature, clearly visible in the \civ\ \lam\
1548.202 line, though present also in the \lam\ 1550.774 line, is
a weak feature near 0 \kms\ (figure~\ref{fig:gd659}). Fitting a
double Gaussian to the co-added \civ\ lines, velocities
of\linebreak 36.74 \err\ 2.56 \kms\ (primary), and -2.97 \err\
3.00 \kms\ (secondary) are obtained. The secondary component is
weak (with an equivalent width of 6 m\AA\ compared to 36 m\AA\ for
the photospheric component), and is of comparable strength to
scatter in the adjacent continuum regions. Although an {\sl
F}-test indicates that a dual Gaussian fit is preferred over a
single component at the 94\% confidence interval, the similarity
between this feature and the natural scatter in the data suggests
that this is simply noise. However, until high resolution \stis\
data for this region can rule out the existence of shifted
components in the \civ\ lines, we tentatively include GD 659 among
the stars with possible circumstellar features.
\\

\subsubsection{Stars with no clear circumstellar features at the resolution of current data}

\begin{figure}
\centering
\rotatebox{270}{\resizebox{!}{9cm}{\includegraphics{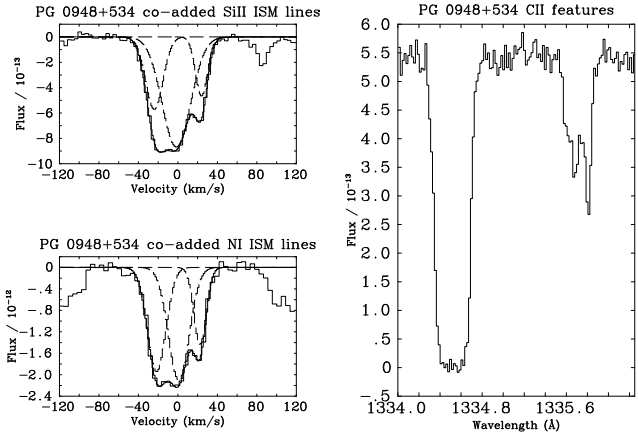}}
} \caption[Interstellar lines of \siii, \NI\ and \cii\ in
REJ~0948+534]{Upper left: co-added interstellar Si {\scriptsize
II} lines in REJ~0948+534, with three-component Gaussian fit.
Lower left: analogous plot for N {\scriptsize I}. Right:
interstellar C {\scriptsize II} lines at \lam\lam
1334.5323,1335.7076.} \label{fig:rej0948}
\end{figure}

{\em PG 0948+534}\\ PG 0948+534 is the hottest star in this
sample. Strong ISM lines are observed, many of which are
saturated, and at least two velocity components are present.
However, \NI\ and \siii\ features are most accurately described by
three components, with self consistent velocities: the primary
component lies at -0.26 \err\ 1.26 \kms, with the secondary and
tertiary components at -22.8 \err\ 1.2 and 22.1 \err\ 2.4 \kms.
These are most clearly revealed by co-addition of the \NI\ and
\siii\ features, as shown in figure~\ref{fig:rej0948}.

The photospheric lines, including the resonance doublets of \civ\
and \siiv, exhibit narrow, symmetrical profiles and are apparently
devoid of any shifted components, defining \vphot\ = -14.25 \err\
0.22 \kms. A remarkably strong, multi-component \cii\ 1335.7076
\AA\ feature is observed, also shown in figure~\ref{fig:rej0948}.
The velocity components match those of other ISM lines, although
the -23 \kms\ feature is very weak, manifesting itself as a
broadening on the blue side of the line. Excited \siii\
transitions such as \lam\lam 1265.002,1309.276,1533.431 are not
observed (\citet{Hol:1995b} used the presence of these lines in
the white dwarf CD~-38\deg\ 10980 to infer the existence of a
circumstellar cloud around the star). No evidence of highly
ionised non-photospheric material is found in PG 0948+534.\\

{\em REJ 2214-492}\\ Weighted average line velocities indicate
\vism\ = -1.72 $\pm\ 0.51$ \kms, and \vphot\ = 33.49 $\pm\ 0.45$
km s\per. These values compare well with those of HBS, who find
\vism\ = -0.71 \err\ 0.88 and \vphot\ = 33.91 \err\ 0.47 \kms\
respectively. However, a significant difference exists between the
velocity of each line in the \civ\ doublet, with \lam\ 1548.202 at
30.5 \err\ 2.1 \kms, and \lam\ 1550.774 at 40.4 \err\ 2.7 \kms, if
each of the two lines is assumed to be made up of only one
absorption feature.

\begin{figure}
\centering
\rotatebox{270}{\resizebox{!}{8cm}{\includegraphics{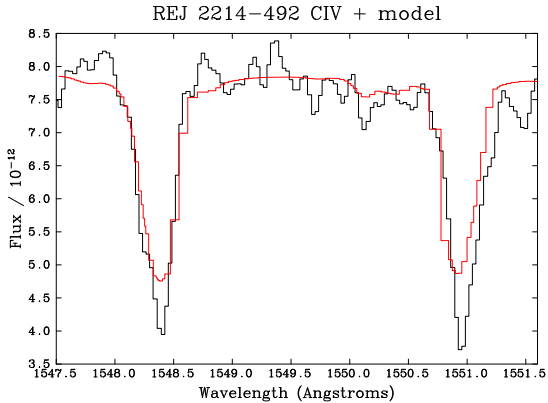}}
} \caption[Observed \civ\ doublet in REJ~2214-492 compared to a
synthetic spectrum]{Observed C {\scriptsize IV} doublet in
REJ~2214-492 (black) compared to a synthetic spectrum with N(C)/ =
$10^{-6}$ N(H), smoothed to the resolution of \iue.}
\label{fig:rej2214}
\end{figure}

Visual inspection of the \civ\ doublet reveals slight asymmetry,
particularly in the \lam\ 1548.202 line, where a dual fit was
found to be superior to the single Gaussian at the 99.9\%\
confidence level, with velocity components at 5.37 and 38.36 \kms,
and equivalent widths of 31.1 and 99.3 m\AA\ respectively. A dual
Gaussian fit to the \lam\ 1550.774 line produced a less obvious
improvement (at the 90\%\ confidence level) with components at
36.2 and 80.6 \kms, and equivalent widths of 137.6 and 14.8 m\AA\
respectively. The primary Gaussian components of the doublet thus
lie at velocities more consistent with each other and the overall
photospheric value. The status of putative non-photospheric
contributions in the \civ\ doublet is less certain; although the 5
\kms\ feature at 1548 \AA\ appears to provide a good match to
observation, the lack of a corresponding feature at 1550 \AA\
prevents confirmation of its reality. No evidence was found for
multiplicity in other photospheric lines. Line profiles were
compared with those from a model spectrum, produced using the
\tlu\ and \syn\ codes, and adopting the heavy element abundances
determined by~\citet{Bar:2000} (with N(C)/N(H) = 1.0 $\times
10^{-6}$). After smoothing the model to the resolution of \iue, no
significant differences were apparent in the shapes of model and
observed \civ\ lines (figure~\ref{fig:rej2214}), casting further
doubt on the presence of shifted features in this star.\\

{\em REJ~0623-371}\\ The photospheric velocity determined here
(\vphot\ = 41.15 \err0.56 \kms) agrees with that of HBS, though
the current value for \vism\ is somewhat lower than given by HBS
(16.40 \err 0.70 \kms\, c.f. 19.48 \err 0.85 \kms\ respectively);
the principal source of this difference lies in a more precise
determination of the \cii\ line velocity. HBS estimate the
velocity of this line to be 14.07 \err 3.24 \kms, compared to the
new value of 13.33 \err 1.14 \kms.

\begin{figure}
\centering
\rotatebox{270}{\resizebox{!}{9cm}{\includegraphics{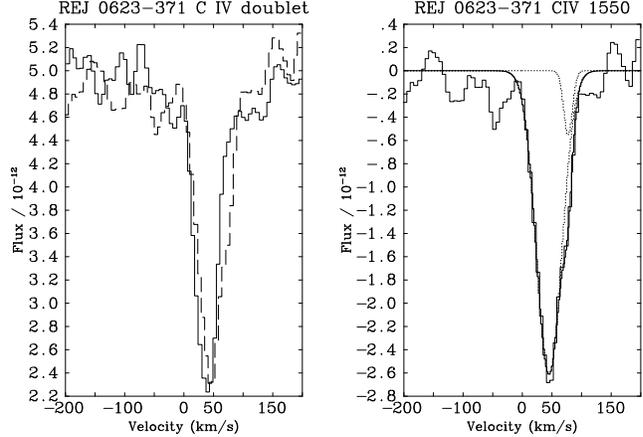}}
} \caption{Left: lines of the C {\scriptsize IV} doublet in REJ
0623-371, in velocity space, demonstrating the difference in
velocity of the two lines (\lam 1548 = solid histogram, \lam 1550
= dashed histogram). Right: dual Gaussian fit to the 1550 \AA\ C
{\scriptsize IV} line (dashed curves = component Gaussians, solid
curve = summed Gaussian.} \label{fig:rej0623}
\end{figure}

No compelling evidence exists for the presence of shifted
components in the spectrum of REJ 0623-371, but as in the case of
REJ 2214-492, a significant difference is observed in the velocity
of the lines in the \civ\ doublet (1548 \AA\ = 38.6 \err 2.3 \kms,
1550 \AA\ = 47.6 \err 2.2 \kms). In contrast, the lines of the
\nv\ and \siiv\ resonance doublets, which are of comparable
equivalent width, agree within the estimated error. Determining
the reality of any features in the doublet is complicated by
considerable absorption in the continuum of this extremely metal
rich DA, though by restricting Gaussian fits to the region below
this structure, useful comparisons between the level of agreement
found with single and double Gaussian profiles, may be obtained.
Using this method, a dual Gaussian fit is preferred to a single
feature only at the 88.8\% level for the 1548 \AA\ line. However,
results for the 1550 \AA\ line are less ambiguous, suggesting a
dual fit at 98.9\%. The resulting Gaussians have velocities of
45.1 and 77.3 \kms, and equivalent widths of 128.5 and 10.8 m\AA\
respectively (figure~\ref{fig:rej0623}). Thus, while the quality
of the available data is insufficient to prove the existence of
circumstellar features in the star, the results of this analysis
provide some justification for proposing repeat observations at a
higher signal-to-noise ratio (\sn) and spectral resolution.\\

{\em REJ 2334-471}\\ Values measured for \vism\ and \vphot\ are in
good agreement with those obtained by HBS. The relatively poor
quality of the \iue\ data precludes unambiguous identification of
any non-photospheric features, particularly in the case of the
\civ\ lines. There is no evidence of multiplicity in the \nv\
doublet, although a dual Gaussian fit to the 1242 \AA\ line, with
components at 19.7 and 43.51 \kms, produces a fit which is
preferred over a single feature at the 93\% confidence level.

It is therefore intriguing that each line in the \siiv\ doublet
(\lam \lam 1393.755, 1402.777) is fitted reasonably well (above
the 95\% confidence interval when compared to a single feature) by
double Gaussian profiles. Each line can be described by a double
Gaussian with V$_1$ = 34.00 \err 0.82 \kms, EW$_1$ = 48.85 \err
0.95 m\AA, and V$_2$ = 54.64 \err 1.58 \kms, EW$_2$ = 22.74 \err
3.25 m\AA. Neither of the Gaussian velocities are in agreement
with the average photospheric value, although a considerable
spread in individual photospheric velocity measurements is
observed, so that the discrepancy cannot be used to infer the
absence of such features. Spectra with improved \sn\ are required
to confirm or disprove the existence of circumstellar features in
this star.\\

{\em GD 246}\\ \iue\ and \stis\ data were available for this star.
Photospheric and ISM line velocities measured from the \iue\ data
are in good agreement with those of HBS. The photospheric velocity
estimated from \stis\ data differs from the \iue\ value by 1 \kms,
although this is within the \iue\ error bounds. The \stis\ ~ISM
velocity (-5.78 \err 0.12 \kms) is marginally outside the \iue\
estimate (-7.87 \err 1.00 \kms). \iue\ and \stis\ data clearly
show the \civ\ doublet lines as singular, and at the photospheric
velocity. \stis\ data shows the \siiv\ \lam\ 1393.755 line as
being devoid of any secondary components, while a sharp feature
(only one bin, or 0.02 \AA\ in width) is observed on the redward
edge of the \lam\ 1402.777 line. Although \iue\ data also hint at
a broadening on this side of the line, unconvincing dual Gaussian
fits, and the extreme narrowness of the extra feature in \stis\
data, suggest that this is due to noise, and thus GD 246 shows no
clear evidence of circumstellar material.\\

{\em PG 1123+189}\\ In the photometric study of hot white dwarfs
by~\citet{Gre:2000}, this object is one of those listed as having
a significant IR excess, suggesting the possibility of a low mass
companion to the star.

The \stis\ spectrum for this object is limited in coverage (1163
-- 1361 \AA), and has relatively low \sn. However, many
interstellar lines are visible in the spectrum, and a value of
\vism\ = -0.67 \err\ 0.06 \kms\ is obtained. Although photospheric
features are difficult to distinguish in the data, by co-adding
the \nv\ lines at 1238 and 1242 \AA\ with those of \niv\ between
1250 and 1336 \AA\ in velocity space, a single absorption feature
is clearly visible, suggesting a value of \vphot\ = 12.55 \err\
0.53 \kms. The quality of these data is insufficient to confirm or
rule out the presence of non-photospheric features with
confidence.\\

{\em HZ 43}\\ A well studied object, HZ 43 is a member of the
group of white dwarfs which can be adequately modeled with an
atmosphere devoid of any heavy elements. Several ISM lines are
observed, leading to an estimate of \vism\ which agrees with that
of HBS. Co-addition of the spectrum at the wavelengths of the
major N, C, Ni and Si lines fails to reveal any photospheric
features. Similarly, co-addition, in velocity space, at the
wavelengths of the excited Si transitions (\lam\lam 1264.738,
1265.0020, 1309.2758 and 1533.4312) also shows no new features.\\

{\em REJ 1032+532}\\ This object is the subject of a comprehensive
study by~\citet{Hol:1999a,Hol:1999b}. In the current work, the
measured value of the primary ISM features, \vism\ = 0.84 \err
0.21 \kms, agrees, within error, with that of~\citet{Hol:1999b}. A
previously noted secondary component to the \siii\ \lam\lam
1193.2897, 1260.4221 and 1526.7065 lines is found to have a
velocity of -30.43 \err 1.39 \kms, also in agreement with the
value quoted by~\citeauthor{Hol:1999b}. A value of \vphot\ = 38.16
\err 0.40 \kms\ is determined for the photospheric velocity. The
excited \siii\ lines found around some stars possessing
circumstellar clouds~\citep{Hol:1995b} are absent, and in none of
the photospheric lines is any compelling evidence found for the
existence of secondary components.\\

{\em PG 1057+719}\\ This object (alternative ID REJ 1100+713) is
also included in the photometric study of white dwarfs
by~\citet{Gre:2000}, with no significant IR excess being detected.
It belongs to the low opacity metal poor class which includes the
majority of DA white dwarfs.~\citet{Hol:1997b} presented a study
of this star and REJ 1614-085 (see below). Their results revealed
no signs of circumstellar features.

The current work confirms the results of~\citeauthor{Hol:1997b},
revealing no shifted features in the \ghrs\ data. Co-addition of
the ISM lines results in \vism\ = -2.89 \err 0.69 \kms. As
expected for a low EUV opacity object, no significant photospheric
lines are observed. To detect any signs of photospheric features,
a series of 10 \AA\ - wide sections were extracted from the data,
each centred on the rest wavelength of one of the lines of the
\nv, \civ\ and \siiv\ resonance doublets. The sections were then
transformed into velocity space, and co-added. The presence of
barely-detectable quantities of N, C and Si might then be expected
to produce a noticeable reduction in continuum level around the
photospheric velocity. The co-added data does indeed reveal a
feature, with a velocity of 75.35 \err\ 2.59 \kms, consistent with
the value of \vphot\ = 76.1 \err 3 \kms\ determined from Balmer
line fitting. However, as indicated by ~\citeauthor{Hol:1997b},
weak individual features found near the expected positions of
these lines show a considerable spread in velocity, casting doubt
on their authenticity.\\

{\em GD 394}\\ GD 394 is photometrically variable in the EUV,
though no signs of spectroscopic variation have been detected. In
contrast to REJ 1614-085, GD 394 has an extreme overabundance of
Si compared with model predictions~\citep{Hol:1997b}.
~\citet{Dup:2000} note that this extreme Si abundance, and the
observed EUV variability, give a unique status to GD
394.~\citeauthor{Dup:2000} present spectroscopic and timing
analyses of GD 394, which suggest the presence of a large EUV-dark
spot on the surface of the star, sharing the stellar rotation
period of 1.150 days. Episodic accretion is proposed as the source
of this spot, with a magnetic field directing material onto the
magnetic poles. No evidence exists for the presence of a magnetic
field in GD 394, though only upper limits can currently be placed
on the strength of any such field. GD 394 appears to be an
isolated star, and hence no obvious candidate exists for the
source of accreted material, other than the immediate stellar
neighbourhood.

Early results suggested that the velocity of \siiii\ and \siiv\
lines differed considerably from the established radial velocity,
and that these lines were therefore of a circumstellar
origin~\citep{Bru:1983}. It was also suggested that the absence of
any observable \siii\ features in \iue\ data, which models
predicted would be present, represented further evidence for the
non-photospheric nature of the heavy elements.
However,~\citet{Bar:1996b} demonstrated that the previous radial
velocity measurement, obtained from the Balmer lines, was in
error, and a revised value more consistent with the velocities
found for the Si features was obtained. By using the latest NLTE
models available at the time,~\citet{Bar:1996b} showed that the
predicted abundance of photospheric Si would yield line strengths
within the noise of the \iue\ spectrum, and hence the absence of
\siii\ features did not require a non-photospheric solution.
~\citet{Cha:2000} report the first firm detection of heavy
elements other than Si in the spectrum of GD 394, with spectra
from the Far Ultraviolet Spectroscopic Explorer ({\em FUSE})
showing lines of \feiii\ and \pv\ in the photosphere, as well as a
large number of \siiii\ and \siiv\ lines.

Values of \vism\ = -7.28 \err\ 1.42 and \vphot\ = 28.75 \err\ 0.91
are obtained in this work for the velocity of ISM and photospheric
features respectively. As reported by~\citet{Hol:1997b}, the
inventory of photospheric lines is dominated by \siiii, with
additional contributions from the \siiv\
doublet.~\citeauthor{Hol:1997b} also observe photospheric \aliii\
lines at \lam\lam1854.7159 and 1862.7900. However, \civ\ and \nv\
are conspicuous only by their absence; co-addition of the data for
these wavelengths reveals no trace of features.

No record exists, in recent literature, of circumstellar features
in the spectrum. The \iue\ \ data reveal no sign of \civ\ or \nv,
and \ghrs\ data are available only for the ranges 1290 --
1325~\AA\ and 1383 -- 1419~\AA, which do not include these ions.
However, many Si lines are visible in the higher resolution \ghrs\
data, and neither individual features or the co-added profiles of
all visible Si lines show any form of asymmetry or other qualities
which may indicate the presence of circumstellar features. GD 394
is therefore one of the few stars in this survey for which it can
be conclusively stated that no shifted features exist, at the
resolution of currently available data.\\

{\em GD 153}\\ Another example of a star which may be modeled with
a pure-H atmosphere, GD 153 is a frequently observed standard
star. No obvious photospheric features are observed in the \iue\
spectrum of this star, although several ISM lines are recorded,
indicating a value of \vism\ = -8.42 \err\ 2.68 \kms. This
velocity agrees with that obtained by HBS.\\

{\em EG 102}\\ \citet{Hol:1997c} showed that \mgii\ (\lam\ 4481)
and excited \siii\ was present in the optical spectrum of this
cool star. Since radiative levitation calculations predict that
much higher temperatures are required before Mg can be suspended
in the atmosphere, observable quantities of this species in EG 102
were interpreted as indicative of ongoing accretion, either from a
low mass companion (for which no evidence was found) or from a
diffuse interstellar cloud. Later, HBS analysed the \iue\ ~NEWSIPS
spectrum of EG 102 and noted the presence of \alii\ and \aliii\
lines at the photospheric velocity. This observation was also
interpreted as being a product of ongoing accretion. Photospheric
and ISM velocities determined for EG 102 in this work are in good
agreement with those measured by HBS.

Data provided by ~\citet{Cha:1995b} indicate an abundance of N(Al)
= $10^{-9}$ N(H) for a star with \lg\ and \teff\ similar to those
of EG 102. The Al abundance in EG 102 has been estimated by
fitting data with a pure H model spectrum into which Al has been
added, using the code \syn. After smoothing the output spectrum to
the resolution of \iue, abundances of order N(Al) = $1 \rightarrow
5 \times 10^{-8}$ N(H) are obtained. These figures are
significantly higher than those suggested
by~\citeauthor{Cha:1995a}, but the work of~\citeauthor{Cha:1995a}
does not adopt the self-consistent approach found in more recent
work (e.g.~\citet{Dre:2001,Sch:2001}, discussed in section 1), and
a failure to accurately predict the Al abundance in EG 102 cannot
be used to infer the presence of accretion processes in the star.

We note that~\citet{Zuc:1998} find significant quantities of Ca in
EG 102\footnote{Listed under the alternative name G238-44 in that
work.} (Ca/H = $2.5 \times 10^{-7}$, {EW = 29m\AA}\ for the \lam\
3933 Ca K line) using high resolution optical echelle spectra
obtained with the Keck telescope; current theories of radiative
levitation do not predict the Ca ion to be present in such a cool
DA.~\citeauthor{Zuc:1998} also find significant quantities of Ca
in both of the close white dwarf/red dwarf pairs in their survey,
and suggest that the presence of heavy ions such as Ca in cool DA
stars may be attributed to binarity. To date, no companion to EG
102 has been observed.\\

{\em Wolf 1346}\\ The presence of Si in the photosphere of Wolf
1346 was revealed by~\citet{Bru:1982}, but later questioned
by~\citet{Ven:1991}, who noted discrepancies between the velocity
of \siii\ lines and the ground-based photospheric velocity, and
found that abundances determined from the \siii\ lines were
inconsistent with the non-detection of \siiii\ in the \iue\
spectra of this star -- observations more suggestive of a
circumstellar origin. The problem was resolved
by~\citet{Hol:1996}, who revised the photospheric velocity and
used the advanced non-LTE code \tlu\ to derive a Si abundance 0.5
dex lower than that of~\citeauthor{Ven:1991}, confirming the
photospheric nature of the Si lines.

The velocities of features normally associated with the ISM appear
to fall into two groups, at -16.19 \err\ 0.10 \kms\ and -7.67
\err\ 3.07 \kms. When considering the velocities of {\sl isolated}
features, these groups appear to contain separate species (\oi,
\cii\ and \SII\ at -16 \kms, and \siii\ at -7 \kms). However, this
segregation breaks down when those interstellar \siii\ features
which are blended with photospheric features, are included. For
example, the ISM component of the \lam\ 1260.4221 \siii\ line,
lies at -16.73 \kms.

The photospheric lines (which are limited to \siii\ and \siiii)
show a considerable spread in velocity (from 19 to 33 \kms), with
the weighted mean being \vphot\ = 24.32 \err\ 1.41 \kms. Given
this range of velocities, the reality of the two ISM groupings is
questionable. Data from HBS also show this grouping, though only
four ISM lines are recorded (compared to ten in the current
study). HBS treat these velocities as belonging to the same group,
and give a weighted mean ISM velocity of -14.85 \err\ 1.50 \kms.
For the current work, the analogous value is \vism\ = -15.38 \err\
0.95 \kms. Thus, both \vism\ and \vphot\ are found to agree with
the values presented by HBS.

\begin{figure}
\centering
\rotatebox{270}{\resizebox{!}{8.5cm}{\includegraphics{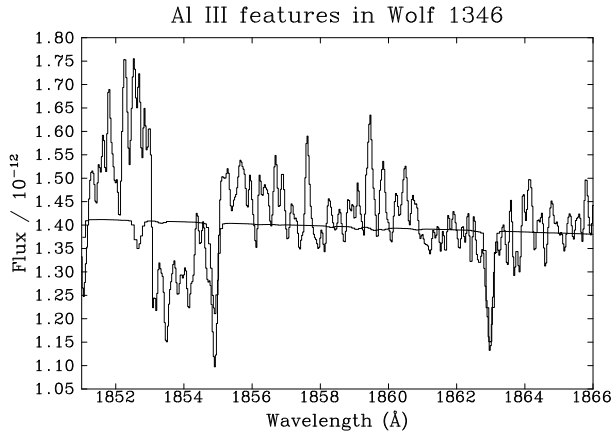}}
} \caption{\lam\ 1854, 1862 \AA\ Al {\scriptsize III} lines in
Wolf 1346, with model spectrum superimposed (darker line). The Al
abundance inferred from the 1862 \AA\ line is approximately 2.2
$\times 10^{-9}$ N(H). Structure in the continuum level around the
\lam\ 1854 line complicated abundance measurements using this
feature.}\label{fig:wolf1346}
\end{figure}

Lines of \aliii\ are clearly visible at the photospheric velocity.
These features are not noted in previous studies of Wolf 1346
(\citet{Hol:1996}, HBS). To estimate the Al abundance, the \lam\
1854, 1862 lines were reproduced by adding quantities of Al to a
spectrum generated from a pure H+He non-LTE model (N(He) =
10$^{-8}$ N(H)), and the output smoothed to the resolution of
\iue\ ($\sim 0.2$ \AA\ FWHM). An abundance of N(Al) = $2.2\times
10^{-9}$ N(H) was found to match the observed 1862 \AA\ line, and
is close to the value of $\sim 1 \times 10^{-9}$ N(H) implied by
the results of~\citeauthor{Cha:1995b}. A greater abundance (N(Al)
= $6.0\times 10^{-9}$ N(H)) was required to match the 1854 \AA\
feature; however, the region around this line exhibits unusual
structure, possibly due to instrumental effects. The models used
to determine these abundances are {\em not} stratified, and hence
these values are likely to be revised when suitable stratified
models become available. The Al lines and synthetic spectrum are
illustrated in figure~\ref{fig:wolf1346}.

\section{Discussion}\label{sec:discussion}
Of the twenty three stars considered in this survey, four were
previously known to possess features from highly ionised species
at non-photospheric velocities: Feige~24, REJ~0457-281, G191-B2B
and REJ~1614-085. Four new DA white dwarfs may now be added to the
list of those exhibiting unambiguous, highly ionised components at
non-photospheric velocities: REJ~1738+665, Ton 021, REJ~0558-373,
and WD 2218+706. A fifth object, REJ~2156-546, shows features
which may also be interpreted as non-photospheric, although data
of improved \sn\ are required to confirm this result. A weak
blueshifted component in GD~659 is also suggested, though this is
exceedingly faint, comparable with the structure of adjacent
noise, and hence regarded as a tenuous identification until
further data become available. Table~\ref{tab:species} lists those
ions appearing as circumstellar features. Observations of these
features are restricted to resonance transitions, and are most
common in the \civ\ \lam\lam 1548.202, 1550.774 doublet (the
resonance lines are most strongly coupled to the stellar radiation
field, and hence are more susceptible to radiative levitation).
The velocities of interstellar, photospheric and non-photospheric
features (if detected) have been presented in
table~\ref{tab:measured}.

\begin{table}
\begin{center}
\begin{tabular}{l l} \hline \hline
\multicolumn{1}{c}{Star} & Species  \\ \hline %
REJ 1738+665 & {\bf \civ, } {\bf \siiv, } \nv, \ov. \\ %
Ton 021 & {\bf \civ, } \siiv. \\ %
REJ 0558-373 & {\bf \civ. }\\ %
WD 2218+706 & {\bf \civ, } {\bf \siiv. } \\ %
Feige 24 & {\bf \civ.} \\ %
G191-B2B & {\bf \civ, \siiv$^*$.} \\ %
REJ 0457-281 & {\bf \civ, \siiv,} \nv. \\ %
REJ 2156-546 & \civ. \\ %
REJ 1614-085 & {\bf \civ, \siiv.} \\ %
GD 659 & \civ. \\ \hline %
\multicolumn{2}{l}{\scriptsize $^*$ Detected by~\citet{Hol:2002}.}
\end{tabular}
\caption[Ionic species contributing to circumstellar
features]{Ionic species contributing to observed or suspected
circumstellar features. Ions responsible for principal features
(i.e. those which are clearly identifiable without the need for
co-addition) are indicated in bold; other species should be
regarded as tentative identifications.} \label{tab:species}
\end{center}
\end{table}

Of the eleven \iue\ spectra considered in this study, only one
(REJ 0457-281) shows signs of highly ionised, non-photospheric
components. In contrast, six out of the twelve available \stis\
spectra reveal such features (with a further detection in one of
the three available \ghrs\ spectra). The low number of detections
in \iue\ data is unsurprising given the low signal-to-noise ratio
of the instrument, and the fact that its resolution (0.08 \AA\ at
1400 \AA) is equivalent to a velocity of 17 \kms\ (compared to
$\sim$ 3.2 \kms\ with the \stis\ ~E140M grating, and $\sim$ 1.28
\kms\ in the E140H configuration). The non-detection of
circumstellar components in \iue\ data therefore places an upper
limit on the velocity of any shifted features, rather than proving
absence.~\stis\ ~data may yet reveal non-photospheric features in
these objects.

The need for consistently high resolution data with adequate \sn,
covering {\em all} stars in follow-up studies is clear. Only the
highest resolution \stis\ data were able to show secondary
features in the ISM lines of G191-B2B, revealing a possible
connection between the circumstellar components and the local ISM.
These features were not resolved in the \stis\ ~E140M data.
However, all but one of the \stis\ spectra included in this survey
were acquired with the E140M grating (PG1123+189 was observed in
the higher resolution E140H mode, but the data do not cover the
\civ\ or \siiv\ resonance lines). Hence, more stars in the current
sample may possess highly ionised non-photospheric features, at
velocity differentials too small for the existing data to resolve.
Alternatively, some stars may possess very weak circumstellar
features which are hidden within the noise of existing data. Only
when \stis\ data (ideally from the E140H configuration and
covering the important resonance lines) are available for all
stars in this sample, will a more comprehensive study of the
distribution of circumstellar features in white dwarfs be
possible.

\subsection{Influence of position and intervening ISM column}
\subsubsection{Spatial distribution}
\begin{figure}
\centering \vspace{-0mm}
\rotatebox{270}{\resizebox{!}{7.5cm}{\includegraphics[bb =120 150
430 740,clip]{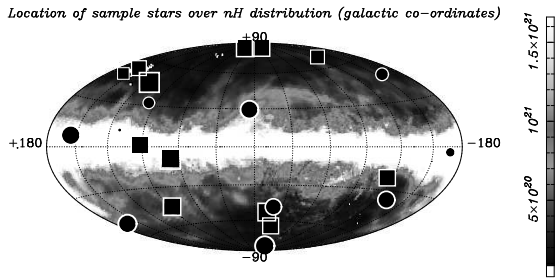}}} \vspace{-0mm}\caption{The
distribution of survey stars in galactic coordinates. Circles
indicate stars in which circumstellar features are observed, with
all other objects plotted as squares. Symbol size indicates
distance (larger = closer). Objects are plotted over the galactic
H {\scriptsize I} map of~\citet{Dic:1990} (units are
cm$^{-2}$).}\label{fig:distribution}
\end{figure}

The distribution of survey stars is depicted in
figure~\ref{fig:distribution}; no correlation between the presence
of circumstellar features and the position of objects on the sky
is apparent. Since the sample stars encompass a relatively wide
range of distances (between approximately 14 and 436 pc), the
absence of such a positional dependence is not surprising.

\subsubsection{ISM N(\hi) column density} Line-of-sight ISM N(\hi) column densities
may be estimated by fitting models to the observed ISM Ly-$\alpha$
profile (after removing the stellar contribution to line+continuum
using a stellar model), as demonstrated by~\citet{Bar:1994b}
and~\citet{Hol:1999b}. Alternatively, the \hi\ column density may
be determined from \EUVE\ data using the strong continuum
absorption below the 912 \AA\ Lyman edge. In this case, the column
density is included as a free parameter in model spectra, which
are fitted to the data using $\chi^{2}$ reduction techniques
(e.g.~\citealp{Bar:1999a,Hol:1999b}). These methods provide the
{\em average} column density, and are insensitive to ``clumping''
along the line of sight. As an extension to this simplification,
it may be assumed that, {\sl ceteris paribus}, a more distant star
along the same or similar line of sight will be observed through
an intervening column, N(\hi), of greater density. The figure of
interest is therefore the volume density n(\hi) along the ISM
column, where, in the general case,
\begin{equation}
N\left(\mbox{\hi}\right)=\int_{0}^{s}n\left(\mbox{\hi}\right) ds\
,
\end{equation}
and where $s$ is the distance to the star. For a homogeneous
column, $n$(\hi) $\approx N$(\hi)/$s$.

Previously determined column densities are available for many of
the sample stars, while approximations may be made for others
using the synthesis maps of~\citet{Frisch:1983}, and the contour
maps of~\citet{Par:1984}. These columns are listed in
table~\ref{tab:columns}, showing that a relationship between
circumstellar features, and the average density of interstellar
material along the line of sight, is not observed. For example,
both REJ~1738+665 and REJ~0457-281 show circumstellar features,
yet their line of sight ISM volume densities are significantly
lower than objects without such features.

\begin{table*}
\caption{H {\scriptsize I} column and volume densities for the
survey stars, obtained from a variety of previous studies. Data
from~\citet{Frisch:1983} and~\citet{Par:1984} should be
interpreted as broad estimations only. No data available for
REJ~0948+534 or EG~102.}
\begin{center}
\begin{tabular}{l r r l} \hline \hline
\multicolumn{1}{c}{Star} & \multicolumn{1}{c}{N(\hi) (cm$^{-2}$)}
& \multicolumn{1}{c}{n(\hi) (cm$^{-3}$)} &
\multicolumn{1}{c}{Reference} \\ \hline REJ 1738+665$^{a,*}$ & $<
5.0 \times 10^{18}$ & $< 6.7 \times 10^{-3}$ & \citet{Bar:1994}\\
Ton 021$^{*}$ & $1\rightarrow 5 \times 10^{20}$ & $1.5\rightarrow
7.5 \times 10^{-1}$& \citet{Frisch:1983}
\\REJ 0558-373$^{*}$ & $1\rightarrow 5 \times 10^{20}$ & $1.1\rightarrow 5.5 \times 10^{-1}$& \citet{Frisch:1983}\\
 REJ 2214-492 & $5.8 \times 10^{18}$ & $2.73 \times 10^{-2}$ & \citet{Wol:1998} \\
REJ 0623-371 & $5.0 \times 10^{18}$ &$1.67 \times 10^{-2}$
&\citet{Wol:1998}
\\ WD 2218+706$^{*}$ & $ > 10^{21}$ &$> 7.4 \times 10^{-1}$ & \citet{Frisch:1983}
\\ Feige 24$^{*}$ & $3.25 \times 10^{18}$ &$1.35 \times 10^{-2}$ & \citet{Wol:1998} \\
REJ 2334-471 & $8.5 \times 10^{18}$ &$2.65 \times 10^{-2}$ &
\citet{Wol:1998}
\\ G191-B2B$^{*}$ & $2.1 \pm 0.1 \times 10^{18}$ & $ 1.4 \times
10^{-2}$ & \citet{Bar:1999a} \\ GD 246 &  $> 1.51 \times 10^{18}$
&$> 6.8 \times 10^{-3}$ &\citet{Fru:1994} \\ REJ 0457-281$^{*}$ &
$(1.3 \pm 0.7) \times 10^{17}$ & $3.9 \times 10^{-4}$ &
\citet{Bar:1994b} \\ PG 1123+189 & $1.10 \times 10^{19}$ & $2.4
\times 10^{-2}$ & \citet{Hol:1999c}
\\ HZ 43 & $9.3 \pm 0.1 \times 10^{17}$ & $4.2 \times 10^{-3}$ &
\citet{Bar:1995a} \\ REJ 1032+532 & $4.2 \times 10^{18}$ & $1.1
\times 10^{-2}$ & \citet{Hol:1999c} \\ REJ 2156-546$^{*}$ & $4.1
\times 10^{18}$ & $1.0 \times 10^{-2}$ & \citet{Hol:1999c} \\ PG
1057+719 & $2.75 \times 10^{19}$&$2.17 \times 10^{-2}$ &
\citet{Wol:1998}
\\ REJ 1614-085$^{a,*}$ & $\approx 5 \times 10^{20}$ &$ \approx 1.89$ &
\citet{Par:1984}
\\ GD 394 & $4.4 \times 10^{18}$ & $2.5 \times 10^{-2}$ &
\citet{Bar:1996b}
\\ GD 153 & $6.03 \times 10^{17}$ &$2.68 \times 10^{-3}$&\citet{Fru:1994} \\ GD 659$^{*}$ &
$3.2 \times 10^{18}$ & $2.0 \times 10^{-2}$ & \citet{Hol:1999c} \\
Wolf 1346 & $ 1.10 \times 10^{18}$ &$2.55 \times 10^{-2}$ &
\citet{Fru:1994}
\\ \hline \multicolumn{2}{l}{~$^{*}${\footnotesize Circumstellar
features observed or suggested.}}
 \\ \multicolumn{2}{l}{~$^{a}${\footnotesize Total column
(N$_{H}$) quoted}} \\
\end{tabular}
\label{tab:columns}
\end{center}
\end{table*}

This result is not surprising.~\citet{Dup:1983} find that for a DA
white dwarf with \teff~=~60,000K and \lg~=~8.0, the Str\"{o}mgren
radius ranges from 0.07 pc (for n(H)=10$^{2}$ cm$^{-3}$) to 30.8
pc\linebreak (n(H)=0.01 cm$^{-3}$). Although their work is now
dated (the heavy element features in white dwarf spectra are
attributed to the ionisation of circumstellar material), these
figures still provide a useful order of magnitude estimate for the
sphere of influence of the white dwarf. It could also be argued
that the quoted Str\"{o}mgren radii represent upper limits, since
the extra opacity from photospheric heavy elements would be
expected to reduce the intensity of the radiation field at
specific wavelengths. In either case, the Str\"{o}mgren radius is
typically a small fraction of the distance to the star, and since
the distribution of material along the line of sight is unlikely
to be homogeneous, the observed average value of n(\hi) may not
reflect conditions within the Str\"{o}mgren sphere.

\subsubsection{The velocity of circumstellar features and the
ISM} For a white dwarf which is ionising nearby interstellar
material, highly ionised non-photospheric components may be
observed at velocities similar to those of the intervening ISM.
Depending on the relative velocities of star and ISM, this
mechanism will produce both red- and blue-shifted features.

\begin{figure} \vspace{25mm}
\rotatebox{0}{\resizebox{!}{6.4cm}{\includegraphics{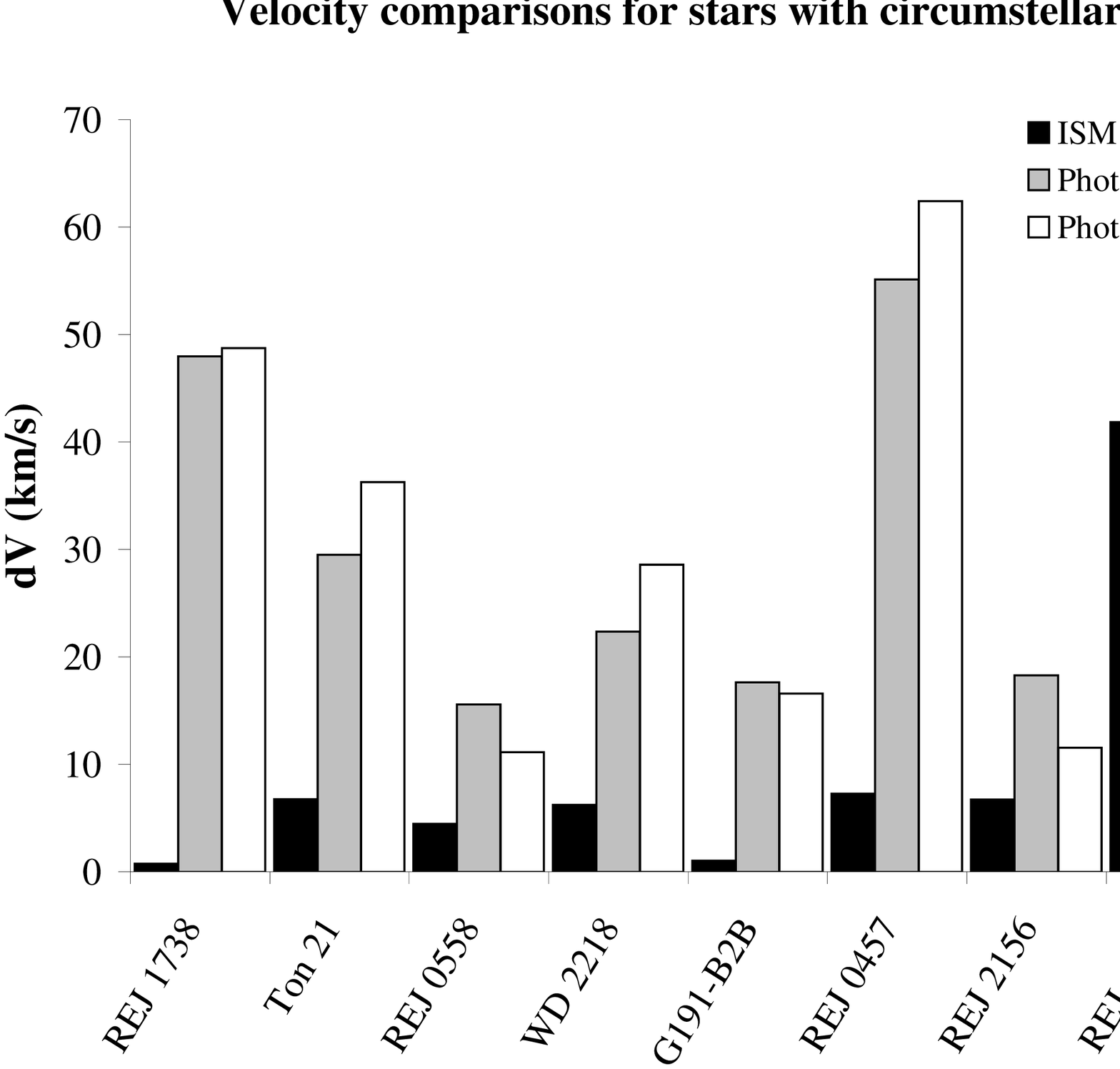}}}
\vspace{-30mm}\caption{Comparison between photospheric, ISM and
circumstellar velocities in stars showing highly ionised
non-photospheric features (Feige 24 omitted since \vcirc\ and
\vphot\ vary with orbital position, while \vism\ is stationary).}
\label{fig:dv}
\end{figure}

The majority of stars possessing circumstellar components show
little agreement between \vcirc\ and \vism. This result is
unsurprising, since observed ISM absorption features may be blends
of several unresolved ISM components at the resolution of the
current data, and hence the dominant component (typically the
LISM) may not be that in which the star is immersed. Nevertheless,
it is useful to consider the residual value of $\mid$\vism\ --
\vcirc $\mid$ (figure~\ref{fig:dv}). These data show that in the
majority of objects possessing circumstellar features, the
difference in velocity between these and the primary interstellar
cloud is typically 10 \kms\ or less (exceptions being REJ 1614-085
and GD 659). The most interesting cases are found in REJ 1738+665
and G191-B2B, where excellent agreement is found between \vism\
and \vcirc, suggesting that the shifted features arise in a
Str\"omgren sphere of material belonging to the primary identified
ISM component.

More detailed investigations are required before this hypothesis
can be confirmed, and a detailed analysis of the ion populations
and spectral characteristics of such a system is essential. The
case of G191-B2B also acts as a caution against
over-interpretation of these results; no correlation between
\vism\ and \vcirc\ would have been recognised if not for the
higher resolution data discussed by~\citet{Sah:1999}. Similar
correlations with hitherto undetected ISM components may be found
in future studies based on E140H grating spectra. Despite the
relatively narrow wavelength coverage available with this
instrument, the current results justify a comprehensive program of
E140H white dwarf observations, possibly tuned to a bandpass
covering the \civ\ resonance doublet, which is most frequently
accompanied by shifted features.

\subsection{Metallicity and mass loss}
\citet{Mac:1992} discusses the interaction between the flow of ISM
material around a white dwarf star, and the weak stellar wind. In
this work, the rate of mass loss, $\dot{M}$, from the white dwarf
is estimated using theory developed by~\citet{Abb:1982}, {\sl
viz.}
\begin{equation}\label{eq:massloss}
\dot{M}\approx 2 \times 10^{-15}
\left(\frac{L}{L_{\odot}}\right)^{2}\frac{Z}{0.02}
M_{\odot}\mbox{~~yr\per},
\end{equation}
where $Z$ is the metallicity, relative to solar abundances, of a
star with luminosity $L$ and mass $M$.

However, the work of~\citeauthor{Abb:1982} is concerned with the
envelopes of O- to G-type stars, and results are found to be most
successful for OB stars. Conversely, the theory does not explain
mass loss rates in Wolf-Rayet stars, which are somewhat different
in structure. Further, since the theory of~\citeauthor{Abb:1982}
is concerned with main sequence objects, the metallicity parameter
is formulated in terms of solar abundances, and cannot be applied
directly to the broad range of heavy element compositions
exhibited in white dwarfs. Hence the use of
equation~\ref{eq:massloss} in the current work is not entirely
appropriate.

Nevertheless, it is interesting to compare the {\sl relative} mass
loss rates of sample objects calculated using
equation~\ref{eq:massloss}. Individual abundances have been
calculated by~\citet{Bar:2000}. These values were determined by
matching observational data to a synthetic spectrum calculated
using \syn, based on a model of appropriate \teff\ and \lg
generated by the non-LTE code \tl. This information is available
for all objects in the sample except PG 0948+534
(table~\ref{tab:abundances}). For each star, the metallicity
parameter, $Z$, is calculated using the expression
\begin{equation}\label{eq:metallicity}
Z=\left(\sum_{z>2}A_{*}(z)\right)\left(\sum_{z>2}A_{\odot}(z)\right)^{-1},
\end{equation}
where $A_{*,\odot}$ is the abundance of the element of atomic
number $z$ relative to hydrogen in the star and in the
Sun\footnote{Using data provided by~\citet{Gre:1998},
$\sum_{z}A(z)_{\odot} \approx $ 1.18E-3, where z=6, 7, 8, 14, 15,
16, 26 and 28, reflecting the elements considered in
table~\ref{tab:abundances}.} respectively. Only ``metals''
(elements heavier than He) are included. Note that although
equation~\ref{eq:metallicity} should be evaluated for all elements
heavier than He, only clearly identifiable species as presented in
table~\ref{tab:abundances}, have been considered here.

\begin{table*}\begin{minipage}{177mm}
\caption{Photospheric heavy element abundances of the surveyed
stars, determined from far-{UV} spectroscopy (no data available
for PG 0948+534). Note that gaps in the table for those stars up
to (and including) REJ~0457-281 are mostly due to the absence of
data for a particular spectral range rather than a true absence of
the element itself. For the remaining, cooler stars, gaps in the
table reflect genuine absence of these species at abundances
detectable by \iue\ or \stis. From~\citet{Bar:2000}.}
\begin{footnotesize}
\begin{tabular}{lllllllll|lll}
\hline \hline Star&C/H&N/H&O/H&Si/H&P/H& S/H& Fe/H & Ni/H & Z & L & $\dot{M}$ \\ \hline %
REJ 1738+665$^{*}$ & 2.0E-8 & 3.0E-7 & 3.0E-7 & 1.0E-6 &  &  & 4.0E-6 & 5.0E-7 & 5.18E-3 & 12.59 & 8.21E-14\\ %
Ton 021$^{*}$ & 5.4E-7 & 8.4E-8 & 1.2E-7 & 2.7E-6 &  & & 1.3E-6 & 6.6E-8 & 4.07E-3 & 10.19 & 4.22E-14 \\ %
REJ 0558-373$^{*}$ & 8.0E-7 & 3.0E-7 & 3.0E-6 & 2.0E-6 & 2.0E-8 & 2.5E-7 & 1.0E-5 & 1.5E-6 & 1.51E-2 & 4.61 & 3.22E-14 \\ %
REJ 2214-492 & 1.0E-6 & 7.5E-8 & 9.6E-7 & 7.5E-7 &  & & 1.0E-5 & 1.0E-6 & 1.17E-2 & 9.38 & 1.03E-13 \\ %
REJ 0623-371 & 1.0E-6 & 1.6E-7 & 9.6E-7 & 3.0E-7 & & & 1.0E-5 & 1.0E-6 & 1.14E-2 & 11.69 & 1.55E-13 \\ %
WD 2218+706$^{*}$ & 4.0E-7 & 1.0E-6 & 1.0E-5 & 6.5E-7 & & & 2.0E-5 & 5.0E-7 & 2.76E-2 & 9.64 & 2.56E-13\\ %
Feige 24$^{*}$ & 1.0E-7 & 3.0E-7 & 5.0E-7 & 3.0E-7 & & & 1.0E-5 & 2.0E-6 & 1.12E-2 & 5.86 & 3.84E-14\\ %
REJ 2334-471 & 2.0E-8 & 5.0E-7 &   & 3.0E-7 &  &  & 1.0E-5 & 5.0E-7 & 9.59E-3 & 2.88 & 7.97E-15 \\ %
G191-B2B$^{*}$ & 4.0E-7 & 1.6E-7 & 9.6E-7 & 3.0E-7 & 2.5E-8 & 3.2E-7 & 1.0E-5 & 5.0E-7 & 1.07E-2 & 3.16 & 1.07E-14 \\ %
GD 246 & & & & 1.0E-7 & & & & & 8.47E-5 & 2.23 & 4.21E-17  \\ %
REJ 0457-281$^{*}$ & 4.0E-7 & 1.6E-7 & 9.6E-7 & 1.0E-7 & 2.5E-8 & & 1.0E-5 & 5.0E-7 & 1.03E-2 & 1.85 & 3.53E-15 \\ %
PG 1123+189 &  &  &  &  &  &  &  & & ---- & 2.75 & ---- \\ %
HZ 43 & &   &   &   &   &   &   &  & ---- & 1.49 & ---- \\ %
REJ 1032+532 & 4.6E-7 & 5.0E-5 &   & 5.6E-8 &  &  &   & & 4.28E-2 & 1.03 & 4.52E-15  \\ %
REJ 2156-546$^{*}$ & & &   &   &  &  &   & & ---- & 1.06 & ---- \\ %
PG 1057+719 &   & & & &   &  &   & & ---- & 0.64 & ----\\ %
REJ 1614-085$^{*}$ & 4.8E-7 & 2.5E-4 &   & 1.0E-8 &  &  &   & & 2.12E-1 & 0.43 & 3.90E-15  \\ %
GD 394 &   &   &   & 8.0E-6 &  &  &  & & 6.77E-3 & 0.45 & 1.39E-16 \\ %
GD 153 &   &   &   &   &  &  &   & & ---- & 0.56 & ----  \\ %
GD 659$^{*}$ & 2.0E-7 & 6.3E-4 &   & 1.6E-8 &  & & & & 5.34E-1 & 0.24 & 3.01E-15  \\ %
EG 102 &   &   &   & 1.0E-7 &  & &   & & 8.47E-5 & 0.03 & 6.54E-21  \\ %
Wolf 1346 &   &   & & 3.2E-8 &  &  & & & 2.71E-5 & 0.03 & 1.91E-21 \\ \hline %
\multicolumn{2}{l}{~$^{*}${\footnotesize Circumstellar
features observed}} \\
\end{tabular}
\label{tab:abundances}
\end{footnotesize}
\end{minipage}
\end{table*}

Stellar luminosities quoted in table~\ref{tab:abundances} have
been calculated from the familiar expression

\begin{equation}\label{eq:luminosity}
L_{*}=10^{0.4\{M_{\odot}-M_{*}\}}L_{\odot} \ ,
\end{equation}
where \Msun\ represents the absolute bolometric magnitude of the
Sun (+4.7), or the star (calculated using the stellar models
of~\citealp{Woo:1995}).

\begin{figure}
\vspace{25mm} \centering
\rotatebox{0}{\resizebox{!}{6cm}{\includegraphics{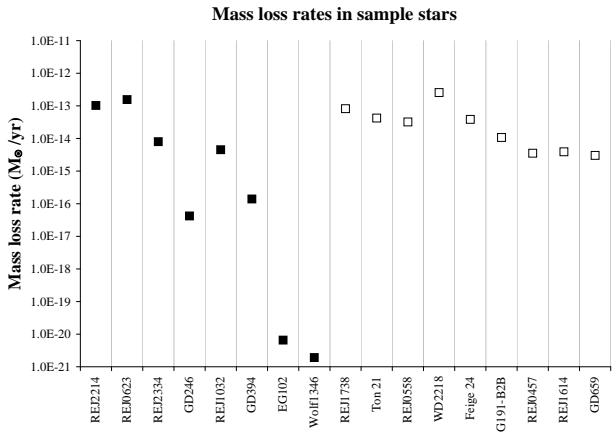}}}
\vspace{-30mm}\caption{Calculated mass loss rates for those sample
stars containing metals, as defined in table~\ref{tab:abundances}.
{\em Filled squares}: stars with no observed circumstellar
features. {\em Open squares}: stars exhibiting circumstellar
features. Objects for which no metals are present or for which
data are unavailable have been omitted.} \label{fig:massloss}
\end{figure}

No clear correlation is found between $\dot{M}$ and the presence
of circumstellar features, although loss rates extend to lower
values for objects without these features
(figure~\ref{fig:massloss}). The lowest values are found in the
coolest stars, as expected given the dependence of
equation~\ref{eq:massloss} on luminosity and metallicity. However,
GD 246 (\teff = 53,700K) is significantly hotter than REJ 1614-085
(\teff\ = 38,500K) despite having an appreciably lower calculated
loss rate. While the subset of stars which exhibit
non-photospheric components lacks any object with $\dot{M} < 3
\times 10^{-15} M_{\odot}$ yr\per, (compared to a minimum value of
$1.9 \times 10^{-21} M_{\odot}$ yr\per\ for those with no
circumstellar components), the number of objects in this survey is
insufficient to determine the authenticity of a lower limit to the
mass loss rates in stars showing highly ionised, non-photospheric
components.

\subsection{Gravitational Redshift}
The apparent velocity of absorption features formed in the white
dwarf atmosphere will be affected by the radial velocity of the
star, and by gravitational redshifting. The velocity change due to
gravitation will be lower in features which are formed in material
further from the stellar surface (e.g. a circumstellar cloud), and
will be effectively zero for a cloud with a sufficiently large
inner radius. The gravitational redshift at the stellar surface
therefore defines a range of velocities, with respect to the
apparent photospheric value, at which highly ionised
non-photospheric features {\em may} be attributed to material
residing within the gravitational well of the star.

Gravitational redshifts can be measured directly in binary systems
such as Feige 24~\citep{Dup:1982,Ven:1994}, but for isolated
systems, the velocity component due to gravitation redshift, \vgr,
can be estimated using the standard formula,
\begin{equation}
\frac{\lambda_{obs}-\lambda_{rest}}{\lambda_{rest}} =
\frac{v_{\hbox{\tiny grav}}}{c}=\frac{GM}{c^2R} ,
\end{equation}
where $G$ is the universal constant of gravitation, $M$ is the
stellar mass, $R$ is the radius (of the star, or of the
circumstellar cloud), $c$ is the speed of light, and
$\lambda_{obs,rest}$ are the observed and rest wavelengths of the
absorption line, respectively. Substituting values for the
physical constants, the expression for \vgr\ becomes simply
\begin{equation}\label{eq:vgr}
v_{\hbox{\tiny grav}}\approx 6.36 \times 10^{2} \frac{M_*}{R_*} \
\  \hbox{m s}^{-1} ,
\end{equation}
where solar units are to be used for the white dwarf mass and
radius. In the work which follows, estimates using this expression
are used to identify features which could arise from material
residing well within the gravitational potential of an object.

Values for \vgr, calculated using equation~\ref{eq:vgr}, are
included in table~\ref{tab:measured}. The presence of material in
the gravitational potential well of a star provides no explanation
for objects in which material appears at a velocity redshifted
with respect to the photospheric value, as in the case of WD
2218+706, Feige~24, REJ~2156-546, and REJ~1614-085. The mechanism
may also be ruled out for three further stars, REJ~1738+665, Ton
021 and REJ~0457-281, in which the calculated gravitational
redshift is substantially lower than required to explain the
velocity of non-photospheric components.

For the three remaining stars, the gravitational redshift is
comparable to the velocity difference between photospheric and
circumstellar components. However, if the calculated values of
\vgr\ are assumed to be reasonable, only circumstellar features in
REJ~0558-373 may be explained by the presence of matter inside the
gravitational well (at approximately 5 stellar radii from the
surface). In the cases of G191-B2B and GD~659, the
non-photospheric components differ from \vphot\ by an amount equal
to, or slightly greater than, the value of \vgr. This suggests
that the material lies at a radius greater than that at which
gravitational redshifting produces observable changes in line
velocity. It is therefore apparent that, with the possible
exception of REJ~0558-373, blueshifted features in the spectra of
objects in this study cannot be explained by the presence of
material within the gravitational potential well of the star.

\subsection{Non-photospheric material and its relation to
planetary nebul\ae} The existence of an old planetary nebula (PN)
around two of the survey stars has already been addressed in the
cases of REJ 1738+665 and WD 2218+706. Mass loss and the
production of a PN as a white dwarf progenitor leaves the AGB, are
relatively well accepted processes (though the precise details are
certainly not~\citep{Lan:1994}). It is therefore pertinent to ask
whether the presence of non-photospheric, highly ionised features
around other DA stars at the hotter end of the sample (i.e. those
for which the nebula material may not be completely dispersed), is
consistent with material from a (now ancient) PN around the star.

Expansion velocities for PNe are available in the literature
(e.g.~\citealp{Wei:1989}). The study by~\citet{Nap:1995} is
particularly relevant, since it deals specifically with old
planetaries, with central stars in the advanced stages of
transition into the white dwarf area of the H-R diagram. Since
even the hottest stars in the current study are more highly
evolved than those covered by~\citeauthor{Nap:1995}, average
temperatures are cooler, and any nebular material presumably of
lower density and more widely dispersed. Direct detection of
nebular emission around these stars is therefore difficult,
excepting the two cases previously highlighted. Further, in the
case of the older, cooler stars in the current work (e.g. REJ
2156-546 and REJ 1614-085), the planetary nebula must have
dispersed long ago, and is therefore unlikely to offer an
explanation for the existence of highly ionised, non-photospheric
features. Nevertheless, the possibility that some circumstellar
features are of this origin may be assessed by comparison with the
expansion velocities noted by~\citeauthor{Nap:1995}.

\begin{figure}
\vspace{25mm}
\rotatebox{0}{\resizebox{!}{6cm}{\includegraphics{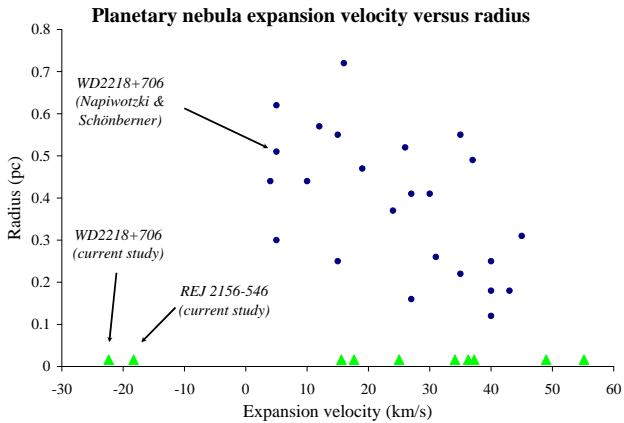}}}
\vspace{-30mm}\caption{Circles: Nebula expansion velocity \vexp\
plotted against nebula radius, for stars studied
by~\citet{Nap:1995}. Triangles: \vphot - \vcirc\ for stars in the
current work (no values for radius are available).}
\label{fig:pneb}
\end{figure}

Figure~\ref{fig:pneb} shows the distribution of expansion velocity
(in this case plotted against nebula radius) for the objects
investigated by~\citeauthor{Nap:1995}. No data are available for
the radius of any PNe which may surround stars in the current
survey; however, an analogous velocity can be derived in the form
of the difference between the velocity of circumstellar and
photospheric features, as listed in table~\ref{tab:measured}. Note
that WD 2218+706 (DeHt5) is common to both studies, and in this
case the true value of \vexp, as determined
by~\citeauthor{Nap:1995}, is considerably different to the {\em
negative} velocity, relative to the photospheric value, found in
the current work. It is therefore clear that the non-photospheric
features of WD 2218+706 discussed in the current study are not
produced by the observed planetary nebula (though the nebula
cannot be discounted as a {\em source} of this material).

We note that the observation of features which are redshifted with
respect to the photosphere is not incompatible with the nebula
hypothesis; for example,~\citet{Twe:1994b} discuss the planetary
nebula Sh 2-174 and the white dwarf GD 561 observed on one edge of
this nebula. The objects are found at similar distances, and
[\oiii] emission in the nebula is located immediately adjacent to
the white dwarf. These observations, the statistical improbability
of GD 561 being an isolated hot white dwarf which happens to be
wandering through the nebula, and the difficulty in explaining the
existence of a small nebula other than being of PN origin, confirm
that GD 561 is indeed the source of the planetary nebula. The
distinctly non-spherical morphology of Sh 2-174 is far from unique
in studies of old PNe (\citeauthor{Twe:1994b} and references
therein). Gross asymmetries are attributed to interaction between
the nebular material and the surrounding ISM through which it
moves. Given the location of GD 561 relative to the central region
of Sh 2-174, it is clear that the PN material may produce spectral
features which are either blue- or red-shifted with respect to the
photosphere, depending on the angle from which the system is
viewed.

For all but two of the stars in this work, \hbox{\vphot - \vcirc}
is of a similar order of magnitude as the expansion velocity
typical of old PNe. Two stars (WD 2218+706 and REJ 2156-546) are
shown with the negative values of \vcirc\ discussed above - but
the absolute values, $\mid$\vphot - \vcirc $\mid$, are consistent
with the remainder of the sample. In contrast, radiatively driven
winds from the surface of a white dwarf should be of a similar
order to the stellar escape velocity, i.e. $\sim$ 1000 km s\per\
or more~\citep{Mac:1992}. These results may indicate some form of
link between the origin of the non-photospheric absorbing matter,
and the old, dispersed PN material surrounding these stars.

The results of this comparison suggest that further work on the
link between highly ionised non-photospheric features, and
planetary nebul\ae\ is justified. Although this hypothesis appears
to contradict the suggestion that shifted features may be related
to interstellar material within the Str\"omgren sphere of a white
dwarf, {\em both} mechanisms may operate in different objects,
with Str\"omgren spheres as the dominant source for cooler
objects. Correlating typical PNe densities with the column
densities derived from curve of growth analysis would provide
further evidence of a relationship between these two apparently
separate entities. This comparison is complicated by the
considerable difficulty in estimating the masses of PNe, arising
from uncertainties in the distances to nebul\ae, and from the
ongoing debate as to whether these objects are typically
ionisation or mass bounded.

\section{Summary}

We have described the detection and interpretation of highly
ionised absorption features at non-photospheric velocities, in
high resolution UV spectra of hot DA white dwarfs. These features
may be indicative of accretion or mass loss in white dwarfs --
processes which may explain the non-equilibrium abundances,
compared with the predictions of radiative levitation theory,
observed in many objects. Four of the stars in the sample were
previously known to show non-photospheric features: Feige 24, REJ
0457-281, G191-B2B and REJ 1614-085. This work has revealed at
least four new objects with multiple components in one or more of
the principal resonance lines: REJ 1738+665, Ton 021, REJ 0558-373
and WD 2218+706. A fifth object, REJ 2156-546 also shows some
evidence of multiple components, though further observations will
be required for their reality to be confirmed.

Several possible mechanisms for the formation of these features
have been discussed. The presence of material within the
gravitational potential well of a white dwarf is found to be an
unsatisfactory explanation for the production of these features.
Predicted mass loss rates based on the luminosity and metallicity
of stars show no correlation with the presence of shifted
features. However, these mass loss rates are calculated using
theories developed for main sequence stars, and may be
inappropriate for application to highly evolved objects. Further,
the quantification of metallicity is a highly subjective
measurement, and is likely to be a major source of uncertainty in
these calculations.

A possible correlation is observed between the velocity of shifted
features and that of the ISM. This is particularly obvious in REJ
1738+665 and G191-B2B, which show very close matches between
\vism\ and \vcirc. For most of the remaining stars, the difference
between these velocities is less than 10 \kms. Higher resolution
observations are required to detect the presence of multiple ISM
velocity components, which may reveal further correlations -- as
demonstrated by the case of G191-B2B.

An alternative or additional source of shifted features may be
found in planetary nebul\ae. Velocities of shifted features with
respect to the photosphere in this study are found to be entirely
consistent with the expansion velocities  typical of old PNe. By
appealing to the irregular morphology of highly evolved nebul\ae,
both blueshifted and redshifted features may be explained.
Detailed modeling of the interaction between the white dwarf and
surrounding material should determine whether stellar radiation
alone is sufficient to produce the observed ionisation, or whether
additional excitation (perhaps in the form of shock-heating) is
required~\citetext{Napiwotzki, private communication}.

The non-detection of highly ionised non-photospheric features in
many of the stars investigated may indicate their absence, but
equally, may reflect the limited resolution and signal-to-noise
ratio of available data. This is particularly important when
considering non-detections in \iue\ data, where velocity
differentials of less than 17 \kms\ between photospheric and
shifted components are below the resolution limit of the
instrument, and where the \sn\ ratio is inferior to that of more
modern instruments such as \stis. This highlights the importance
of acquiring consistently high resolution data for all stars in
this and future samples. Four (and possibly five) new
identifications of circumstellar features have been made using
medium resolution (E140M) \stis\ ~spectra; these successes
demonstrate the value in high resolution studies of this type, and
justify an extension of the program to include higher resolution
data of improved \sn\ for an expanded sample of objects.

\section{Acknowledgements}
NPB and MAB were supported by PPARC. JBH wishes to acknowledge
support of for this work provided by NASA through grant GO-7296
and AR-9202 from the Space Telescope Science Institute, which is
operated by the Association of Universities for Research in
Astronomy, incorporated under NASA contract NAS 5-26555. We thank
Cherie Miskey (Institute for Astrophysics and Computational
Sciences) for assistance in processing \stis\ data, and Ralf
Napiwotzki (Dr. Remeis-Sternwarte Bamberg Astronomical Institute
of the University of Erlangen-N\"urnberg) for useful discussions
relating to planetary nebul\ae\ and their interaction with white
dwarf stars.

\label{lastpage}

\end{document}